\documentstyle[referee,psfig]{mn}

\title{Interaction between the Intergalactic Medium and Galactic Outflows 
from Dwarf Galaxies}

\author[Murakami and Babul]
{ I.~Murakami$^{1}$ and A.~Babul$^{2}$
\\
$^1$ National Institute for Fusion Science, Toki-shi, 
    Gifu-ken 509-5292, Japan \\
$^2$ Department of Physics \& Astronomy, University of Victoria,
P. O. Box 3055, Victoria, BC  V8W 3P6 Canada \\
}

\begin{document}

\maketitle

\begin{abstract}

We have carried out 2D hydrodynamical simulations in order to study the
interaction between supernova-powered gas outflows from low-mass galaxies
and the local intergalactic medium (IGM).  We are specifically interested
in investigating whether a high pressure IGM, such as that in clusters of
galaxies, can prevent the gas from escaping from the galaxy, as suggested 
by Babul and Rees (1992).  We find that this is indeed the case 
as long as ram pressure effects are negligible.
The interface between the outflow and ambient IGM is demarcated by
a dense expanding shell formed by the gas swept-up by the outflow. A
sufficiently high IGM pressure can bring the shell to a halt well 
before it escapes the galaxy.  Galaxies in 
such high pressure environments are however, more likely than not, to 
be ploughing through the IGM at relatively high velocities. Hence, they
will also be subject to ram pressure, which acts to strip the gas from 
the galaxy.  We have carried out simulations that take into account the 
combined impact of ram pressure and thermal pressure.  We find that
ram pressure deforms the shell into a tail-like structure, fragments 
it into dense clouds and eventually drags the clouds away from the 
galaxy.  The clouds are potential sites of star formation and if
viewed during this transient phase, the galaxy will appear to have a 
low-surface brightness tail much like the galaxies with diffuse comet-like
tail seen in z=1.15 cluster 3C324.  The stars in the tail would, in
time, stream away from the galaxy and become part of the intracluster 
environment.

In contrast, the relatively unhindered outflows in low density, low 
temperature environments can drive the shells of swept-up gas out to large
distances from the galaxy.  Such shells, if they intersect a quasar 
line-of-sight, would give rise to Ly $\alpha$ absorption lines of the kind 
seen in quasar spectra.   In addition, the fact that outflows from 
low-mass galaxies can extend out to distances of 40 kpc or more indicates 
that such galaxies may have played an important role in polluting the 
intergalactic medium with metals.


\end{abstract}

\begin{keywords}
galaxies:dwarf -- inter galactic medium 
\end{keywords}

\section{Introduction}

The study of dwarf galaxies has important implications for our current
understanding of processes governing the formation of galaxies, stars,
and large scale structure in general.  The generally accepted models
for structure formation via hierarchical clustering predict the existence
of a large number of dwarf galaxies and that these galaxies are expected
to be the sites of the earliest star formation 
(see  White \& Frenk 1991).  

Although  dwarf galaxies are the most numerous type of galaxies in the 
nearby universe, their numbers are far fewer than theoretical predictions
\cite{fb}.  
Furthermore, the observations also
seem to suggest that the smallest galaxies are among the youngest rather 
than the oldest,  and this phenomena appears to be independent 
of environment.  For example, many of the Local Group dwarf spheroidals 
and dwarf irregular galaxies appear to have formed a significant fraction 
of their stars in starbursts at $z\la 1$ (see, for example, Van den Bergh 
1994 and Tolstoy 1999).  Deep imaging studies of intermediate redshift 
($0.4\la z \la 1.0$) clusters
(e.g.~Dressler et al.~1994; Couch et al.~1994, 1998) 
have found that most of the 
blue galaxies responsible for the Butcher-Oemler effect are small late-type
spirals or irregular galaxies.  It has been suggested that these galaxies
are starburst dwarf galaxies (Koo et al.~1997) that eventually fade away
or star-forming remnants of `harrassed' small galaxies that will eventually 
evolve into the cluster dwarf spheroidal population (Moore et al.~1996; 
Moore, Lake, \& Katz 1998).
Similarly, detailed analyses of the faint blue galaxies that numerically 
dominate the field galaxy population at intermediate redshifts
($0.4\la z \la 1.0$) 
indicates that a significant fraction of the very faint blue galaxies are 
small (and therefore, dwarf) galaxies that are actively forming stars at 
a relatively recent epoch (see Babul \& Ferguson 1996 and references therein; 
Campos 1997; Driver \& Fernandez-Soto 1998; Fioc \& Rocca-Volmerange 1999).

The discrepancy between the theoretical expectation that small galaxies ought
to be among the oldest and the observational evidence to the contrary
can either be due to the fact that hierarchical clustering model does not 
provide a complete description of structure formation particularly on the 
scales of interest, or that the astrophysics underlying the formation of dwarfs 
galaxies is not well understood, with the latter being the more accepted
of the two possibilities.  For example, it has long been recognized that dwarf 
galaxies are rather fragile systems that their formation and subsequent 
evolution, in fact their very character,
is likely to be strongly affected by both internal and external conditions.  
The `galaxy harrassment' model of Moore et al. (1996; 1998), 
the models advocating catastrophic mass loss following 
supernovae explosions
(Larson 1974, Saito 1979,  Vader 1986, Dekel \& Silk 1986), and models that 
argue that star formation in these systems is strongly modulated by the 
internal and external UV radiation field 
(e.g.  Babul \& Rees 1992; Efstathiou 1992; 
Kepner, Babul \& Spergel 1997; Norman 
\& Spaans 1997; Spaans \& Norman 1997; Corbelli, Galli \& Palla 1997) are
all based on the fragility of the dwarf systems.

The `galaxy harrassment' mechanism of Moore et al. (1996; 1998) operates
only in high density enviroments.  Based on results of numerical simulations,
Moore et al. have proposed that multiple, high-speed encounters between
small disk galaxies and the large massive galaxies in the cluster enviroment
can cause multiple starbursts  in, as well as significant mass loss from,
the lower mass galaxies.  The interactions would rearrange the internal 
structure of the smaller galaxies, converting them into much more resilient, 
compact dwarf 
elliptical-like galaxies. Moore et al. also suggested that the continued 
harrassment of the debris tails --- material torn from the original galaxies 
--- will create tidal shocks that will promote further condensations
and formation of dwarf galaxies.  In this scenario, the most numerous
class of galaxies in clusters have formed at moderate redshifts.

With regard to the impact of UV radiation on the star formation in dwarf
galaxies, Babul \& Rees (1992) and Efstathiou (1992) have discussed in 
detail how an ionizing flux of an intense metagalactic UV background, such
as that established by quasars and early starbursts, can  prevent
the gas in halos with shallow potential wells from cooling and forming 
stars until $z \la 1$, in spite of the fact that the halos themselves
may have formed at some earlier epoch, by initially keeping the gas ionized 
and latter, by suppressing the formation of molecular hydrogen, the only 
coolant available to metal-poor gas (see Kepner, Babul \& Spergel 1997). 
However, analytic calculations (Rees 1986; Ikeuchi 1986) and subsequent
numerical studies (Katz, Weinberg \& Hernquist 1996; Thoul \& Weinberg 
1996; Quinn, Katz, \& Efstathiou 1996; 
Navarro \& Steinmetz 1997; Forcada-Miro 1997) 
show that photoheating most strongly affects halos with circular 
velocities $v_{\rm c} \la 30-35 \rm km \ s^{-1}$.  Larger halos are largely
unaffected.  These limits, however, should be treated with some caution
because  none of these studies considered the impact of UV radiation from 
the very first stars on further star formation.  In addition,
all numerical studies except that of 
Kepner et al. (1997)
ignored the fact that before metal-poor gas could cool and form stars, the 
ambient conditions must allow for the formation of neutral hydrogen,
and even Kepner et al. (1997)
only explored models involving
spherical collapse.   Subsequent studies by Corbelli, Galli \& 
Palla (1997), Norman \& Spaans (1997), and Spaans \& Norman (1997)
have gone much further.  The latter authors, for example, considered
halos with total masses spanning the range $10^8 - 10^{12} M_{\odot}$ and 
found that even if a halo does not get `hung-up' by the background UV 
radition, the star formation in disk-like protogalaxies will initially 
proceed very slowly because the backreaction of star formation on the 
ionization and chemical 
equilibrium can greatly impact upon the abundance of $\rm H_2$.  The tight 
coupling between the radiation field and star formation rate is 
reduced only when  the gas is sufficiently enriched.  Consequently, the
onset of massive starbursts, if any, even in Magellanic-type systems
will be delayed until $z\sim 1$.

Once stars do
form,  there appears to a  consensus, at least among the theorists,
that the dwarf galaxies, due to their low escape velocities, will suffer 
supernova-driven outflows (eg. 
Larson 1974, Saito 1979,  
Vader 1986, Dekel \& Silk 1986).  
Whether or not such outflows
result in catastrophic loss of the interstellar medium is likely to depend
on a variety of factors, such as the energy input in the interstellar medium 
(ISM), the ellipticity of the ISM distribution, etc. 
 (see  De Young \& Heckman 1994).
However, it is 
worth noting that the catastrophic loss of the gas for a galaxy and the
subsequent quenching of the star formation provide  a
natural explanation 
for the low surface brightness and low metal abundances of dwarf galaxies
(eg. Dekel \& Silk 1986,  
De Young \& Gallagher 1990).  It also provides an explanation for 
the apparently rapid disappearance of the 
large numbers of small blue galaxies (low mass starbursting systems)
that dominate the galaxy number counts 
at very faint magnitudes (see Babul \& Rees 1992, 
Babul \& Ferguson 1996).

Although supernova-driven outflow is an important aspect of any theoretical
discussion of dwarf galaxy evolution, there has been, until recently, quite
limited observational evidence for the existence of starburst-driven mass loss.
The most detailed and convincing evidence was first reported by 
Meurer et al. (1992), 
who found a kpc-scale `superbubble' of ionised gas expanding 
at $\sim 100$ km/s in the core of the post-starburst dwarf galaxy NGC1705. 
This has now been followed up by 
Marlowe et al. (1995), 
who have studied a large
sample of starbursting dwarf amorphous galaxies and found evidence for 
outflows in approximately half of them.   Also,
Heckman et al. (1995)
have argued that X-ray emission from dwarf galaxy NGC1569 is a signature 
of starburst-driven outflow.  

Generally, it is  assumed that if the starburst imparts sufficient energy to 
the affected fraction of the ISM to unbind it from the galaxy's gravitational 
potential well, then this material will be ejected from the galaxy.  
The situation is not as simple.  The supernova rate needs to be high enough
so that the remnants percolate the galaxy's interstellar medium in a time 
short compared to the remnant's radiative timescale (see, for example,
Dekel \& Silk  1986).  In addition, Babul \& Rees (1992) 
have noted that whether or not the 
outflow actually escapes from the low mass galaxy also depends on the state
of the local intergalactic medium (IGM).  The thermal pressure due
to the IGM will resist the flow of material out of the galaxy.
They argued that in regions of high pressure, such as in clusters of 
galaxies, the outflow would not be able to expand beyond the extent of 
the galaxy's dark halo and hence, 
will eventually accrete back onto the galaxy, allowing the galaxy to engage 
further star formation .  In regions of low thermal pressure, the outflow would
escape unhindered.  

One consequence of the thermal-confinement picture
is that there ought to be correlations between the properties of dwarf
galaxies and their environment.  For example, the low mass galaxies in 
high pressure environments, by virtue of being able to retain a greater
fraction of their gas, ought to be more luminous and more metal-rich
than comparable galaxies in the field.  In clusters of galaxies, one
would expect the dwarf galaxy population to exhibit a radial luminosity and
metallicity gradients, with the galaxies closer to cluster center being, on
the average, brighter and more metal-rich (Babul \& Rees 1992).  Recent 
observations seem to suggest that such trends do exist.  For example, 
Secker (1996) 
finds significant color gradient in the radial distribution
of dE galaxies in Coma, with redder galaxies tending to be closer to the
cluster center.  Secker interprets the color gradient as indicative of a
metallicity gradient.  Similar radial color gradients have also been
detected in other clusters (O. Lopez-Cruz, private communications).  
Secker, Harris \& Plummer (1997) also find that redder dE galaxies are also
more luminous.   More generally, dwarf elliptical galaxies are found only
in high-density environments such as clusters of galaxies or clustered 
around giant galaxies 
(Vader \& Sandage 1991; Ferguson 1992; Vader \& Chaboyer 1993).  According
to the the thermal confinement scenario, this is because  low mass galaxies
in high density environments are more luminous and have higher surface 
brightness, and are therefore easier to detect.  
Ferguson \& Sandage (1991) and  Vader \& Sandage (1991) 
have argued that the number ratio of early-type dwarfs to 
early-type giant galaxies is correlated with cluster richness with the ratio 
being larger for richer clusters.  The bright nucleated dE galaxies 
have only been sighted in clusters and even in this environment, they tend to
be much more centrally concentrated than the non-nucleated dEs 
(Ferguson \& Binggeli, 1994).   We also be note that if the intrinsic
mass function of the galaxies is steeply rising towards the low mass end, as 
predicted by most hierarchical galaxy formation theories, then
this steep end ought to be easiest to observe in high density envrionments 
since the galaxies of a given mass in such environments will tend to be 
relatively brighter.  Luminosity functions of cluster galaxies with steep 
rise at the low luminosity end have been observed by, 
for example, Trentham (1988).

In this first of a series of studies, we investigate the idea of 
pressure-confined outflows put forth by Babul \& Rees (1992).  
Specifically, we use a 2D hydrodynamical
code to study the influence of the local IGM on outflows from 
spherically symmetric dwarf galaxies, such as
the Local Group dwarf spheroidals and cluster dwarf ellipticals,
 in various different environments.  
Admittedly, spherically
symmetric dwarf galaxies are likely to lose the greatest amount of ISM; 
however, in the present study, we are not so much concerned with the degree 
of mass loss as with the interaction of the outflow with the local 
intergalactic environment.  In addition, dwarf galaxies are not static objects.  
In high-density cluster environments, for example, the dwarfs are likely 
to be ploughing through the local IGM at relatively high velocities.  In 
such cases, they will also be subject to ram pressure and contrary to the 
action of thermal pressure, ram pressure acts to strip the ISM from the galaxy
(Gunn \& Gott 1972; Gisler 1976; Lea \& De Young 1976; Toyama \& Ikeuchi 1980; 
Fabian, Schwarz, \& Forman  1980; Takeda, Nulsen, \& Fabian 1984;
Gaetz, Salpeter, \& Shaviv 1987; Portnoy, Pistinner \& Shaviv 1993;
Balsara, Livio, \& O'Dea 1994).
We also investigate 
the combined impact of ram pressure and thermal pressure under different 
conditions. 
In \S 2, we define our model for the dwarf galaxy and outline the
methods used in this investigation. In \S 3.1, we discuss the
evolution of outflows subject only to thermal pressure 
and in \S 3.2, we consider
the combined impact of thermal and ram pressure. In \S 4, we discuss
the possible observational consequences of the interaction between the outflows
from dwarf galaxies and the surrounding intergalactic medium.  Finally,
we summarize the our findings in \S 5.

\section{Model}

Since we are interested in low-mass galaxies, 
we will adopt as a fiducial galaxy, a
system whose dark halo mass is $M=10^9 M_{\sun}$ and the 
circular velocity characterizing the gravitational potential of the system is 
$V_{\rm c} \approx 18$ km/s.  
For simplicity, we assume that the virialized dark halos have 
spherically symmetric density profiles 
\begin{equation}
\rho(r) \propto 1/(r^2 + r_{\rm c}^2), 
\end{equation}
where $r_{\rm c}$ is the core radius set by the numerical resolution of our 
simulation, which is either $0.1$, $0.2$,  or $0.25$ kpc.
The corresponding one-dimensional velocity dispersion 
of the galaxy is $\sigma\approx 25$ km/s.  
This velocity dispersion is somewhat larger than that of a typical 
Local Group dwarf spheroidal ($\sigma\sim 10$ km/s); on the
other hand, the general characteristics of our system (i.e. low 
mass and shallow potential well) are comparable to those of the 
dwarf ellipticals (Peterson and Caldwell 1993) found in clusters
of galaxies.

We assume a mass of $M_{gas}=6 \times 10^7 
M_{\odot}$ for the gas in the halo. In order to be susceptible to star 
formation, the gas must be at least marginally self-gravitating
(eg. Mathews 1972). 
To satisfy this constraint, we require all the gas
to be concentrated within  a central $1$ kpc of the halo and 
identify the
centrally condensed baryonic system in the halo as the galaxy. All
stellar activity, from star formation to supernova explosions, 
takes place in the galaxy. 

The `natural' rate for star formation in a self-gravitating gas cloud is 
\begin{equation}
\dot{M}_\ast\propto {M_g(t)\over t_{ff}}, 
\end{equation}
where $M_g(t)$ is the instantaneous mass of gas cloud and $t_{ff}$ is the 
free-fall timescale for the cloud (e.g.~Dekel \& Silk 1986).
For the system under consideration, the star formation rate is $\sim
1$ $M_{\sun}\rm  yr^{-1} $.   This rate is almost three orders of magnitude
larger than that the maximum rate adopted by  MacLow \& Ferrara (1998) in
their study.  It is, however, comparable to the star formation rate in
the faint blue galaxies as established from their redshifts and B magnitudes
(Babul \& Rees 1992).  Furthermore, since the star formation is distributed
over a region one kiloparsec in radius, the star formation rate per unit
area of $\sim 1$ $M_{\sun}\rm  yr^{-1} kpc^{-2}$ is comparable to that 
seen in typical starburst galaxies (Lehnert \& Heckman 1996).

Soon after the onset of starburst, the massive stars will go supernova.
If each supernova releases $10^{51}$ ergs, the rate of
total energy released by the supernova explosions into the interstellar
matter is 
\begin{equation}
\dot{E}_{\rm  SN }= 10^{49} \varepsilon_{\rm  SN }
   \dot{M}_* \rm  ergs \ yr^{-1} ,
\end{equation}
where $\varepsilon_{\rm  SN }$ is the number of supernovae per
$100M_{\sun}$ of stars formed, and $\dot{M}_*$ is the star formation
rate in units of $M_{\sun}\rm  yr^{-1} $.  In the solar neighbourhood,
one has roughly one supernova for every 150 $M_{\sun}$ of baryons that
form stars; hence, $\varepsilon_{\rm  SN }\approx 0.67$.  If starbursts
make only high mass stars (see, for example, Rieke et al.~1993),
then $\varepsilon_{\rm SN}$ would be larger.  For present
purposes, we  assume that $\varepsilon_{\rm  SN } \dot{M}_* =1.4$.

Most of the energy released by the supernova will 
radiated away and only a small fraction, $\eta\approx 0.1$, will go towards
heating the ISM 
(see Larson 1974; Dekel \& Silk 1986; Babul \& Rees 1992). 
The total rate of energy input to gas is then
$\dot{E}_{\rm  Heat } = \eta \dot{E}_{\rm  SN }
       \approx 1.4\times 10^{48}\rm  ergs \ yr^{-1} $.
We assume that this energy input lasts for $2 \times 10^7$yr.
The lifetimes of supernova Type II progenitors range from few$\times 10^{6}$--
few$\times 10^{7}$ years, and we assume that the very first generation of 
SN explosions will disrupt the interstellar medium and quench further
star formation.

As we will show in the following section, the energy input described above
is sufficient to generate an outflows from the dwarf galaxy.  This is 
neither surprising  nor the thrust of the present work. Many  previous
works (eg. Larson 1974; Dekel \& Silk 1986; Babul \& Rees 1992) 
have already
made this point.  Here, we are more interested in studying how different
local intergalactic conditions affect the outflow and the eventual fate of
the expelled gas.  Specifically, we focus on the influence of the
thermal and ram pressures engendered by the intergalactic medium.

At the start of the simulation, the IGM gas is uniformly distributed across
the entire simulation volume except in the galaxy, the central 1 kpc region
of the halo.  This distribution is not in equilibrium with the gravitational 
potential of the halo and in the absence of outflows from the galaxy, would 
accrete into the halo.  In our simulations, this accretion is not important.
The associated timescale is comparable to the total duration of a single 
simulation run and is much longer than all the dynamical timescale established
by the outflows.

To study the influence of thermal component of the intergalactic pressure, we 
vary the value of
 $\widetilde{P}_{\rm  IGM }\equiv nT$ from $10^{-2}$ to $10^5$ K \ cm$^{-3}$.
The lower limit corresponds to an IGM of $n\sim 10^{-6}$ cm$^{-3}$ heated to 
the temperature of $T\sim 10^4$ K by photoionisation and the upper
limit corresponds to pressures thought to exist in central regions of
galaxy clusters.  As we have already noted, observations indicate
that starburst dwarf galaxies are found in all type of environments, especially
at intermediate redshifts.

To study the effects of ram pressure, we allow
the galaxy halo to have velocity, with respect to the IGM, ranging
from $V=400$ to $1000$ km s$^{-1}$.  The parameters for the cases discussed
in this paper are summarized in Table 1. 

We study the outflow and its interaction with the IGM using an 
axisymmetric  Euler code 
(Norman \& Winkler 1985;  Yoshioka \& Ikeuchi 1990;
Murakami \& Ikeuchi 1994).  
Following Norman \& Winkler (1985), 
we include artificial viscosity in our simulations in order to treat shocks.
We use the code to 
solve (in cylindrical coordinates) \hfill\break
the continuity equation:  
\begin{equation}
  \frac{\partial \rho}{\partial t} + \nabla \cdot ( \rho \bmath{v})=0,
                                       \label{eq-continuity}
\end{equation}
the Euler equation:  
\begin{equation}
 \rho \left[ \frac{\partial \bmath{v}}{\partial t} +
     (\bmath{v} \cdot \nabla )\bmath{v}\right] =
  - \nabla P - \rho \nabla \Phi,
                                        \label{eq-motion}
\end{equation}
the Poisson equation:  
\begin{equation}
   \nabla^2 \Phi = 4 \pi G  \rho_d,
                                        \label{eq-poisson}
\end{equation}
and the energy equation: 
\begin{equation}
  \frac{\partial}{\partial t}(\rho \epsilon)
  + \nabla \cdot ( \rho \epsilon \bmath{v}) =
  -P \nabla \cdot \bmath{v} + \cal{H-L}.
                                       \label{eq-energy}
\end{equation}
In the above equations, 
$\epsilon$ is the specific internal energy, ${\cal H}$ and 
${\cal L}$ are the heating and cooling rates respectively, 
and $\rho_d$ is the density of dark halo matter.  We ignore the 
self-gravity of the baryonic system.  We also ignore the depletion 
of gas as it forms stars as well as mass input from stellar winds
and supernova explosions.  The rest of the symbols have their
usual meanings.

The heating rate, ${\cal H}$, is zero everywhere except in the galaxy 
itself, where for a brief time, the supernova explosions inject thermal 
energy into the ISM at a rate 
${\cal H} = 3 \dot{E}_{\rm  Heat }/ 4 \pi r_{\rm o}^3$,
where $r_{\rm o}=1$kpc. 
For the cooling rate, we consider two possibilities: A cooling rate for 
gas with primordial abundance (hereafter, referred to as `primordial cooling') 
and a cooling rate for gas with cosmic abundance ($Z=0.017$) determined by 
Allen (1973)
(hereafter, referred to as `metal cooling').  
In order to compute the primordial cooling rate,
we take into account radiative recombination, collisional ionisation,
thermal bremsstrahlung, line emission and dielectric recombination (see
Umemura \& Ikeuchi 1987)
for the gas of $n_H/n_{He}=9$.
For the metal cooling rate, we  use a broken 
power-law fit of the results of Raymond, Cox \& Smith (1976)
who adopted the cosmic abundances (that is, $X=0.73$, $Y=0.25$, and 
$Z=0.017$) tabulated by Allen (1973).
In computing the heating rate due to supernova explosions, we have already
taken into account the fact that most of the energy will be lost via cooling
radiation.  To keep the cooling functions on during this time would mean
that we would be double counting the cooling losses. 
Therefore, the cooling function is switched on after the supernova explosions 
cease.

\section{Results}

We now discuss the results of our simulations.  First, we shall consider 
cases where the outflows from dwarf systems are impeded only by the thermal 
pressure of the IGM.  Subsequently, we shall consider cases where the dwarf 
system is acted upon by thermal as well as ram pressure. 

\subsection{The Impact of Thermal Pressure on the Outflow}

Energy input from supernova explosions causes the interstellar medium
in the galaxy to heat up to $T\sim 10^6$ K.  As a consequence of 
the resulting pressure differential between the gas in the bubble and 
the IGM, the bubble expands and an outflow is established.  At its
maximum, the leading edge of the outflow has a velocity of order 
$200 \rm  km \ s^{-1} $.  
The outflow/expansion represents the conversion of some of 
the SN-injected thermal energy into kinetic energy.  As in all outflow-type 
situations (e.g. see Castor, McCray \& Weaver 1975 
and references therein), 
the interaction between the outflow and the intergalactic medium leads to 
the formation of a dense shell consisting of the swept-up ISM and IGM.  
In Figures \ref{fig-1} -- \ref{fig-4},
we show the evolution of the outflow for models -2M, 0M, 3M 
and 4M (see Table 1).  The results for gas subject to primordial cooling are 
qualitatively similar.

In cases where the IGM temperature is $T=10^4$ K (models -2M and 0M/P), the 
leading edge of the outflow is supersonic and the interface between the outflow 
and the unperturbed IGM is demarcated by a shock.  In models where the IGM 
temperature is $T=10^7$ K, the outflow is always subsonic and, as expected,
the gas ahead of the forming shell is also perturbed.  
In Figure \ref{fig-5},
we show
the radial pressure, density, temperature, and velocity profiles for models 
0P and 0M.  The left and right panels show the profiles for models 0P and 0M
respectively during the expansion phase.  
In Figures \ref{fig-6} and \ref{fig-7},
we show
the expanding (left-hand panels) and contracting (right-hand panels) for
models 4M and 4P, respectively.  During the heating phase, the pressure 
in the bubble exceeds the IGM pressure and the heated gas begins to expand
outward.  At the head of the outflow, the swept-up ISM and IGM begins to form 
a shell.  Once the outflow is established and the heating stops, the gas
in the cavity cools as a result of radiative as well as adiabatic
cooling.  The higher density gas closer to the center of the gas
cools much more rapidly than lower density cavity gas behind the shell
as a result of  efficient radiative cooling.  Eventually, all of the gas
in the cavity cools.  

Once the pressure in the cavity drops below that of the external IGM,
the shell begins to decelerate.  (The sweeping up of the IGM also 
decelerates the shell but in all but the low pressure cases, 
[$\widetilde P_{\rm  IGM } \la 10^2$], the effect is not important.)  
With continued expansion, the pressure in the cavity may fall below that
in the shell and a contact discontinuity forms at the boundary between
the two.  Once the shell velocity becomes comparable to the sound speed 
in the external IGM, the shell velocity stagnates.  If a contact
discontinuity has formed between the cavity gas and the inner boundary
of the shell, then in the shell frame of reference, the inner shell boundary
begins to expand inward into the cavity, led by a shock front which 
halts and thermalizes the outflowing cavity gas and also smoothes out
the sharp pressure gradient.  This inward expansion of the inner shell boundary
has also been noted by Ciardi \& Ferrara (1997).

In high pressure environments, the stagnation radius and the halting radius 
are, for all practical purposes, equivalent as the stagnation point is 
reached just before the shell is stopped.  The inward expansion of the inner 
shell boundary is, however, evident in the right panels (collapsing phase) 
of Figures \ref{fig-6} and \ref{fig-7}.
In low pressure environments, the shell decelerates very 
slowly and there can be a large lag between the time when the shell velocity 
stagnates and when it comes to a halt.  During this time, the outer shell
continues to expand outward while the inner boundary begins to expand
inward, and the shell thickness grows.  For $\widetilde P_{\rm  IGM } 
=10^0$ (Figure \ref{fig-5}),
the stagnation occurs at $R\approx 20$ kpc.

As expected, the main difference between the metal and primordial cooling
cases is the increased efficiency of radiative cooling in the former
case.  This is obvious both from the rapid evolution of the temperature 
profiles,
the thinness of the shell and the  dramatic drop in the temperature of the gas 
remaining in the galaxy.  In fact once the heating stops, a small
cooling flow is established as the gas flows  back towards the potential
minimum.

The maximum radius to which the shell expands depends on the 
IGM pressure.  If the IGM pressure is low ($\widetilde P_{\rm  IGM }\la 10^2$), 
the shell can be driven well beyond the virial radius of the halo and into 
the intergalactic space. If, on the other hand, the IGM pressure is as
high as in the central regions of clusters, the shell expands by a very 
small amount before being halted.  From a physical point of view, the 
evolution of the bubble-shell system is schematically similar to that 
of `superbubbles' as sketched out by 
MacLow, McCray \& Norman (1989) 
although, it should be noted that are differences between the configuration
that they studied and those considered here. One consequence of this
is that the bubble-shell expansion in our simulations is not self-similar.

The simulation results do, however, suggest that the radius of maximum 
expansion is related to the value of the IGM pressure as:
\begin{equation}\label{max-shell-radius}
R_{\rm  max }\approx 
  8.2 \left({\widetilde P_{\rm  IGM }\over 10^{3}}\right)^{-0.36}
\ \ \rm  kpc ,
\end{equation}
with $R_{\rm  max }$ for the metal cooling cases tending to be slightly
smaller than that for the corresponding primordial cooling cases.
In addition, preliminary numerical experiments indicate that 
$R_{\rm  max }$ also depends, albeit weakly in the case of shallow 
potential wells of the kind under consideration in this paper, on the 
depth of the gravitational potential well of the halo in the sense that 
$R_{\rm  max }$ decreases as $V_c$ increases.

During the expansion phase, the amount of intergalactic mass that is 
swept-up by the shell-bubble system is:
\begin{equation}\label{swept-up-mass}
M_{\rm  shell }\approx 6.3\times 10^{6} 
 \left({n_{\rm  IGM }\over 10^{-4} {\rm  cm^{-3} }}
\right) \left(\frac{R}{10 \rm  kpc }\right)^3\ \ \rm  M_\odot.
\end{equation}
In all but cases 0P/M and -2M, the external pressure brings the 
shell-bubble  system to a halt well before the swept-up mass exceeds
the mass originally in the SN-heated bubble.  

For models -2M, the shell continues expanding during the entire course
of the simulation, expanding to distances greater than $40$ kpc.  In this
case, the shell-bubble system would need to expand out to $98$ kpc before
the swept-up mass exceeds the initial mass.  We can safely assume that the 
system will indeed expand out to this radius since $R_{\rm  max }$ for this
configuration is estimated to be $\sim 500$ kpc.  However, we do not 
follow the expansion out to such radii.

For model 0P/M, the bubble-shell system expands beyond $21$ kpc, the
radius at which the swept-up mass equals that initial mass in the bubble.
Thereafter, the leading front of the shell continues to evolve much
like an `Oort snowplow'.  $R=21$ kpc is also roughly the stagnation radius
and therefore, while the outer shell radius continues to expand, the inner
boundary expands inwards as described previously, sweeping through the cavity
in $\sim 7 \times 10^8$ years.  This inward propagating shock is stable 
against Rayleigh-Taylor instabilities because both the density gradient and 
pressure gradient have the same sign.  And in this regard, the inward moving 
shock is very different from the shell collapse that occurs in high pressure 
environments and that we discuss below.

Once the shell is brought to a halt, both gravity and the 
IGM pressure begin to force the system to contract.  The timescale
for the shell to fall back onto the central galaxy, assuming that all the
gas has piled up in the shell and that the shell remains intact during 
the collapse, can be estimated as
\begin{equation} 
\tau_{\rm  crush } \sim 
     \sqrt{ \frac{M_{\rm  bubble }}{4 \pi R_{\rm  max } P_{\rm  IGM }} } 
           \approx 5\times 10^7 
        \left(\frac{\widetilde{P}_{\rm  IGM }}{10^3}\right)^{-0.32} \rm  yrs , 
\label{eq-tinfall} 
\end{equation}
where we have used equation (\ref{max-shell-radius}).  This estimate 
matches the actual collapse time within a factor of $2$.  In computing
$\tau_{\rm  crush }$, we have assumed that in comparison to the pressure force,
the gravitational force can be neglected;  the pressure force exceeds
the gravitational force by two orders of magnitude or more.

The bottom two panels of Figures \ref{fig-3} and \ref{fig-4}
show the collapse of the shell in 
cases 3M and 4M, while the right panels in 
Figures \ref{fig-6} and \ref{fig-7}
show the 
the radial pressure, density, temperature, and velocity profiles for models 
4M and 4P, respectively, at different times during the collapse phase. 
The evolution of the SN-heated gas bubble as well as the shell during the
collapse phase depends sensitively on the efficiency of cooling. In the metal 
cooling case, the
cooling timescale of the gas in the bubble is generally shorter than the
dynamical timescale and gas/shell collapses in a `simple' fashion.  In
the primordial cooling case, however, the cooling timescale is larger
than the dynamical timescale.  As the gas/shell collapses, the gas
is heated and the resulting increase in the pressure causes the shell to
bounce.  (See curves for $t=6.1 \times 10^7$yr and $6.8 \times 10^7$yr 
in the right-hand panels of Figure \ref{fig-7}.) 
The same, though a bit more
pronounced, occurs in simulations with no radiative cooling.

Once the shell starts to collapse, it begins to buckle and deform into
tentacle-like structures.  This deformation is due to initially
small perturbations that are present in the shell being enhanced by
Raleigh-Taylor (R-T) instability.  As discussed by Chevalier (1976)
and others, a pressure-driven flow is R-T unstable if the pressure 
gradient (source of acceleration) and the density gradient 
have opposite signs.  In numerical simulations of supernova 
explosions (e.g. Nagasawa, Nakamura, \& Miyama 1988; Arnett, Fryxell, \& 
Muller 1991; Hachisu et al. 1991), R-T instability manifests as well-defined
mushroom-like features.  This shape is a consequence of the fact
that the structures develop outward.  In present case, the shell
is contracting. The features associated with the R-T instability
are also mushroom-like; however, the heads of the mushrooms develop
inward where there is less volume and hence, merge with each other,
giving rise to tentacle-like structures.  

Whether R-T instability materializes in a numerical simulation
and the extent to which it does depends sensitively on the resolution
of the numerical simulation \cite{afm}.
In Figure \ref{fig-8},
we show the evolution of shell-bubble for model 
3M during the collapse phase.  The two left panels show the results
corresponding to resolution d$r=0.2$ kpc and the two right panels
show the results for exactly the same simulation but with d$r=0.1$ kpc.

In both cases, the shell buckles and tentacle-like features arise
as it is forced to contract.  In the higher resolution case, however,
there are many more tentacle-like features, the shell itself is
thinner and has a greater tendency to fragment forming small clouds.
This is especially true of the lagging sections of shell.  
The presence of clouds and extended tentacles ensures that the collapse 
is not uniform.  The lag between when the first parts of the shell
reach the central region of the galaxy and when the clouds
fall in is $ \sim 1.3 \times 10^7$ yr. Once the shell has fragmented and 
the pressure surrounding the clouds has equilibrated, the clouds are only 
subject to gravitational forces.

\subsection{The Impact of Thermal and Ram Pressure on the Outflow}

As we mentioned in the introduction, galaxies in high-thermal-pressure 
environments such as clusters of galaxies are also moving, often
supersonically, through the intracluster medium.  The intracluster medium
flowing through a  galaxy results in the gas in the galaxy being subjected
to ram pressure forces (in addition to the thermal pressure forces).
The impact of the ram pressure is to strip away the gas in a galaxy, 
eventually denuding the galaxy of its gas content.  Ram pressure stripping
is thought to be the dominant mechanism by which galaxies in cluster
environment lose their gas 
(Gunn \& Gott 1972; Gisler 1976; Lea \& De Young 1976) 
and consequently, a great deal of effort has gone
into trying to understand the process, especially through the use
of two-dimensional hydrodynamic simulations 
(eg. Takeda et al. 1984;
Gaetz et al. 1987; Portnoy et al. 1993;
Balsara et al. 1994 
).  

In the case of galaxies moving through the intracluster medium
at velocities    comparable to or larger than 
the velocity dispersions of typical clusters
($v\sim 1000 \rm  km \ s^{-1} $), the ram pressure forces can equal or exceed 
the thermal pressure forces (we shall continue to denote thermal pressure as 
$P_{\rm  IGM }$): 
%
\begin{equation} \label{eq-ramp}
\frac{P_{\rm  ram }}{P_{\rm  IGM }} \sim \frac{ \rho v^2}{nkT} 
 \sim  \left( \frac{V_{\rm  flow }}{\sigma_{dis} }  \right)^2,
\end{equation}
where $\sigma_{dis}$ is the velocity dispersions of clusters.
Ram pressure effects can have significant
impact on the outflow from the galaxies.  
At the very least, the expanding
gas shell will not be spherically symmetric in spite of the fact that
the outflow is.  Upstream, the expanding shell is subject to both
thermal and ram pressure ($P_{\rm  IGM } +P_{\rm  ram }$) 
while on the downstream
side, the shell is largely unaffected by ram pressure.  Consequently,
the shell should therefore be oval-shaped.  The halting 
distance of the shell should also reflect this asymmetry.
For example, the upstream stopping distance of the shell 
associated with a galaxy moving at $1000$ km/s through a 
$\widetilde{P}_{\rm  IGM }=10^3$ ICM should be comparable to the maximum
expansion radius of the shell associated with a galaxy embedded in
$\widetilde{P}_{\rm  IGM }=10^4$ ICM (see equation \ref{max-shell-radius}).  

The snapshots in Figure \ref{fig-9}
show the effect of ram pressure on the outflow 
from a dwarf galaxy described in the example above.  The galaxy is moving at 
a velocity of $V_{\rm  flow }=1000 \rm  km \ s^{-1} $ 
through a medium characterized by 
$\widetilde{P}_{\rm  IGM }=10^3$ (Model 3MW).  For reasons mentioned above, the 
expanding shell has an oval shape (Figure \ref{fig-9}a). 
Once the upstream expansion
of the shell is halted, ram pressure begins to drag the material in the shell
downstream, distorting the shell (Figure \ref{fig-9}b). 
The downstream side shell is also distorted by the eddying flow towards
the center of potential well.
The shell buckles and begins
to fragment into high density clouds (Figure \ref{fig-9}c,d).
In time, the upstream segment of the shell is 
pushed back into the galaxy 
while the remains of the rest of the shell is distorted
into a hyperboloid-like surface.  The galaxy-shell system resembles a comet,
with the main galaxy forming the head and the shell material being the tail.
The dense clouds are cold and are potential sites of star formation. 
If this was 
to occur, one would expect the resulting galaxy to
have a relative high surface brightness `head' attached to a diffuse tail.

While the evolution of the outflow/shell in models 4PW/4MW are qualitatively 
similar to that in models 3PW/3MW (described above), 
this is not the case when the 
ambient thermal pressure is as high as $\widetilde{P}_{\rm  IGM }=10^5$ (models 
5MW/5PW).  In such circumstances, the combined pressure (ram $+$ thermal) 
greatly exceeds the thermal pressure of the supernova heated gas.  Even
during the heating phase, heated gas is unable to expand upstream.  Instead,
the shell that forms at the interface between the supernova-heated gas and 
ambient intergalactic medium is quickly swept downstream and the galaxy loses
most its gas in $\sim 2\times 10^7$ years (Figure \ref{fig-10}).  
Perpendicular to the flow, the heated gas manages to expand slight 
but the resulting larger cross-section presented to 
the oncoming wind only hastens the sweeping away of the gas.

Thus far, we have been considering the impact of ram pressure on a
dwarf galaxy that is moving at a velocity comparable to what one would
expect of a galaxy in a rich cluster.  We have also considered cases
where the galaxy is moving at somewhat lower velocities, velocities
comparable to what one would expect in poor clusters and galaxy groups.

Specifically, we consider
cases where $\widetilde{P}_{\rm  IGM }=\{10^2,\  10^3\}$ and
$V_{\rm flow}=\{400,\  600\} \rm  km \ s^{-1} $, respectively. 
In these simulations, the 
thermal pressure is not overwhelmed by ram pressure 
and during the expansion phase,
the results are similar to that in the no-wind case except that the shape
of the expanding shell is not quite spherical.  The main effect of the
wind is to gradually drag downstream the more slowly collapsing tentacles and 
clumps. 
Most of the expelled gas manages to collapse back into the central regions 
but it too is eventually dragged out of the halo and carried downstream.

The clouds that form during the crushing and the fragmentation of the shell 
tend to have densities $n_H \geq 0.1 \rm  cm^{-3}$ and sizes 
$R_{\rm  cl }\sim 0.1$--$0.5$ kpc.  
The cloud masses range between 
$M_{\rm  cl }\approx 10^5$--$10^6 M_{\sun}$.  
Once formed, the clouds are accelerated
by the wind.  For a cloud at rest, the timescale for cloud to be accelerated
to the escape velocity of the dwarf galaxy halo by ram pressure is
\begin{eqnarray} 
\tau & \sim &  \frac{v_{\rm  esc } M_{\rm  cl }}{P_{\rm  ram }
 \pi R_{\rm  cl }^2}
        \nonumber \\
 & = &
       2.6 \times 10^7 
          v_{\rm  esc,20 } M_{\rm  cl,5 } V_{\rm  flow,3 }^{-2}
          n_{\rm  IGM,-4 }^{-1} R_{\rm cl,-1}^{-2} 
            \rm  yr ,
\label{tescape}
\end{eqnarray}
where
$ v_{\rm  esc,20 } = v_{\rm  esc }/ 20~\rm  km \ s^{-1} $,
$ M_{\rm  cl,5 } = M_{\rm  cl }/ 10^5~\rm  M_{\sun} $,
$ V_{\rm  flow,3 }= V_{\rm  flow }/ 10^3~\rm  km \ s^{-1} $,
$ n_{\rm  IGM,-4 }= n_{\rm  IGM }/ 10^{-4} \rm  cm^{-3}$, 
and $ R_{\rm cl,-1} =  R_{\rm cl}/0.1~\rm  kpc  $.
The clouds remain in the vicinity of the galaxy (i.e.~within the dark halo
of the galaxy) for approximately $10^8$ years
before they are dragged away.  
The clouds are also subject to thermal evaporation,
which will enhance the stripping rate
\cite{nul}.  

%
%
\section{Discussion and Speculations}

One of the most important results that we draw from our work is that in
cluster environments, the confinement of the supernovae-heated 
outflows from dwarf galaxies is complicated by the effects of ram pressure.
If the ram pressure acting on a dwarf galaxy is much less than the thermal 
pressure of the local intergalactic medium, then, as described by Babul \& 
Rees (1992), the outflow from the galaxy is indeed halted by the thermal 
pressure and the confined gas subsequently falls back onto the 
galaxy, providing fuel for a possible second burst of star formation.  
Otherwise, ram pressure alters the confinement picture in a very significant
fashion.   
Only a small fraction of the expelled gas manages to collapse
back onto the central galaxy; most of it is swept away.  If this is the
case, it is difficult to understand how trends reported, for example,
by Secker (1996) would arise unless there exist two different populations
of dE galaxies, an original population that is centrally concentrated within
the cluster and whose members have lower velocities, and a more extended,
higher velocity population comprising of galaxies that fell into the 
cluster at some latter
time.  The outflows from the former group would be subject to thermal
confinement and one would expect such galaxies to be brighter and redder
on the average, much like the population of nucleated dEs.  The more 
extended `infall' population of dE galaxies, on the other hand, 
are likely to recover a very small fraction of the outflowing gas:  If 
the outflow occurs while the galaxies are outside the cluster, thermal ICM 
pressure there is
too low to effect any confinement.  If the outflow occurs after the galaxies 
fall into the cluster, the ram pressure acting on the galaxies would 
sweep away the bulk of the gas.

In cases where the ram pressure is important, the gas that is swept away 
tends to be concentrated in small clouds --- fragments of the shell that formed
at the interface between the ICM and the outflowing material.  Ordinarily,
transfer of heat from the IGM and into the clouds via conduction would cause 
the clouds to evaporate.  The magnetic fields that permeate cluster 
environments are likely to suppress thermal conduction and therefore, 
the clouds will maintain their integrity.  The
clouds are dense enough to support small episodes of star formation and 
because of their velocities and spatial distribution, we would expect that if
the galaxy was observed while the clouds were forming stars, it would resemble
a comet, with the actual galaxy forming the head and the distributed star 
forming regions tracing out the `comet tail'.  There are quite a few galaxies 
with `comet-like' morphologies in $z=1.15$ cluster 3C324 (Dickinson, private 
communications; Dickinson 1996).  Drawing upon the results of our simulations, 
we speculate that these objects are galaxies whose interstellar medium is 
being stripped away and that some of the stripped material is undergoing 
star formation, giving rise to the diffuse tail-like structures.  This would
suggest the structures should have blue colors, possibly bluer than the colors
of the central galaxy 
although it is difficult to quantify the expected difference
in color between the `head' and the diffuse `tail' because infalling material
collapsing onto the central galaxy may also cause the central galaxy 
to experience a burst of star formation.  

Furthermore, we would argue that the tail-like structures are transient 
features that will eventually disappear.  
As noted above, the ram pressure accelerates the 
clouds to relatively high velocities and, although once formed
the stars are immune to effects of ram pressure, 
they will eventually disperse because of the velocity
imparted at the time formation (in effect, the cloud velocity at the time).  In
cluster environments, where in fact stripping is most likely to occur, the high
velocity stars would give rise to a diffuse population of intraclusters stars.
Such a population of stars have recently been detected in Fornax 
(Theuns \& Warren 1997)
and Virgo (Ferguson, Tanvir \& Von Hippel 1998). 

As already noted, the evolution of the gas streaming out of dwarf 
galaxies in environments where the ICM pressure is low, is very different
from that of dwarfs in hot, high density regions.  The outflow triggered by
the first generations of supernova explosions will give rise to a
mass shell that, for all practical purposes,  expands away from the galaxy
and carries away its gas supply.  The field dwarf galaxies, therefore, 
are likely to experience only one short episode of star formation.

The dense expanding shells and dense clumps will give rise to
Ly $\alpha$  absorption lines in quasar spectra if lines of sight to the
quasars intersect such structures.  Here we consider the profiles of
such absorption lines and the HI column densities associated with the
absorption.  As seen in the simulations, almost all of the galactic gas, 
 $M_{gas}$,  and the swept-up IGM are in the expanding shell. 
When the shell radius, $R_{\rm s}$, is much larger than 1kpc, 
the shell mass is
\begin{eqnarray}   
M_{\rm s} & \simeq & M_{gas} + 4 \pi  \mu m_H n_{\rm IGM} R_{\rm s}^3 / 3 ,
                           \nonumber \\
    & \simeq & M_{gas} ( 1 + 0.06 R_{\rm s,30}^3 n_{\rm IGM,-6}M_{g,7.8}^{-1}),
\end{eqnarray}
where $m_H$ is the mass of the 
hydrogen atom,  $R_{\rm s, 30}=R_{\rm s}/30\;\rm kpc$,
$    M_{g,7.8}=  M_{gas} /6 \times 10^7 M_{\sun}$,
 and 
$n_{\rm IGM, -6}=10^{-6} \rm cm^{-3}$.  For simplicity, we assume that the 
gas has primordial abundance ($n_H/n_{He}=9$) and therefore, $\mu=1.3$.
The swept-up gas mass dominates the shell mass once the shell expands beyond
$R_{\rm s} \sim 77 (M_{g,7.8}/n_{IGM,-6})^{1/3}$kpc.  
If the IGM density is higher,
the swept-up mass can come to dominate sooner.  For 
$n_{\rm IGM}=10^{-4} \rm cm^{-3}$, the IGM density in outer regions of the 
clusters or the mean IGM density of the universe at $z=3.5$,
the swept-up mass becomes
comparable to the galactic gas when $R_{\rm s} \sim 17 M_{g,7.8}^{1/3}$ kpc. 
The average hydrogen density of the shell is then
\begin{eqnarray}
 n_H & = & 0.9 M_{\rm s} /(4 \pi \mu m_H R_{\rm  s }^2 \Delta R_{\rm  s } )
           \rm  cm^{-3}, \nonumber \\
     & \simeq & 1.5 \times 10^{-4}
           M_{g,7.8} 
         R_{\rm  s,30 }^{-2} \Delta R_{\rm s,1}^{-1} \nonumber \\
     &        & \ \ \ \ \ \ \ \ \times 
                   ( 1 + 0.06 R_{\rm s,30}^3 n_{\rm IGM,-6}M_{g,7.8}^{-1})
         \rm cm^{-3},
\end{eqnarray}
where  
$\Delta R_{\rm s}$ is the width of the shell, and
$\Delta R_{\rm s,1}=  \Delta R_{\rm s}/1 \rm kpc$.

If the gas is photoionised by a metagalactic UV flux and is in ionisation
equilibrium, the number density of neutral hydrogen is given by
$
 n_{\rm HI} \simeq n_H^2 \alpha_H /G_H J,
$
where $\alpha_H$ is the hydrogen recombination rate, and
$G_H J_0$ is the photo-ionisation rate (see, for example, Black 1981).   If
$J_0$ is the UV flux at Ly limit 
($J_0=J_{21}\times 10^{-21}\;\rm erg \ s^{-1} cm^{-2}Hz^{-1}sr^{-1}$) 
and the spectrum of the background UV radiation is 
$J_\nu \propto \nu^{-1}$, then the neutral hydrogen number density is
\begin{eqnarray}
n_{\rm HI} & \simeq  &3.1  \times 10^{-9}
          T_4^{-3/4} J_{21}^{-1}
         M_{g,7.8}^2 
R_{\rm s,30}^{-4} \Delta R_{\rm s,1}^{-2} \nonumber \\
     &        & \ \ \ \ \ \  \times
 ( 1 + 0.06 R_{\rm s,30}^3 n_{\rm IGM,-6}M_{g,7.8}^{-1})^2
         \rm cm^{-3},
\end{eqnarray}
where $T_4=T/10^4\;\rm K$ is the gas temperature.  The HI column density 
at zero impact parameter through the shell is
\begin{eqnarray}
N_{\rm HI} &  \simeq & 2 n_{\rm HI} \Delta R_{\rm s} \nonumber \\
       & \simeq & 1.9  \times 10^{13} T_4^{-3/4}J_{21}^{-1}
        M_{g,7.8}^2 
R_{\rm s,30}^{-4} \Delta R_{\rm s,1}^{-1} \nonumber \\
     &        & \ \ \ \ \ \times ( 1 + 0.06 R_{\rm s,30}^3 
        n_{\rm IGM,-6}M_{g,7.8}^{-1})^2
         \rm cm^{-2}.
\label{eq-NHI}
\end{eqnarray}
The column density depends on the shell radius. 
When the IGM density is low, $N_{\rm HI}$ decreases rapidly with increasing 
shell radius.  However, once the swept-up mass becomes comparable to the
galactic mass, the above equation suggests that $N_{\rm HI}$ 
becomes proportional 
to  $R_{\rm s}^2$ but as we have seen in Figure \ref{fig-5}, 
the shell thickness does not remain constant.  
At large radii, the shell thickness too increases as mass is swept-up and 
this increase in the shell width slows the growth of the HI column density.

Figure \ref{fig-11}
shows the HI column density along a line-of-sight that 
has impact parameter, $p$, as a function of time for model -2M.
The radius of the expanding shell is indicated by the upper horizontal 
axis. The radius is defined as the outermost boundary of the shell.   For 
the background UV radiation, we use $J_{21}=1.0$ and neglect any 
self-shielding of the UV flux.
The HI column density through the center grows with time because of the 
cooling inflow of gas.  On the other hand, the maximum value of the 
HI column density for lines with $p >1$kpc
decreases as the shell radius grows.


The shell motion and structure affects the shape of the absorption line. 
A typical line-of-sight intersects a shell in two places and therefore,
one would expect to see two absorption lines separated in velocity space
by $2V_{\rm s}\sqrt{R_{\rm s}^2 - p^2}/R_{\rm s}$ because of the 
expansion of the shell.
When the separation between the two lines is small,
thermal broadening of the lines can, however,  cause the lines to overlap 
and blur, giving the impression of one broad line.  
Alternatively, the 
combination of the two lines can give rise to  double-horn features.  
The absorption profiles that take into account the velocity field and
thermal broadening can be calculated according to Wang (1995).

Figures  \ref{fig-12} and \ref{fig-13}
 show the absorption line profiles for model 
-2M at $t=1.9 \times 10^8$yr (Figure \ref{fig-12})
and for model 0M at $t=2.9 \times 10^8$yr (Figure \ref{fig-13}).
The impact parameters are  the same as in
Figure~11.  The double-horn line profiles caused by the two absorption line
features are readily observed, particularly
at low impact parameters.  As the impact parameter is increased and 
the component of the expansion velocity along the line-of-sight becomes 
smaller, the separation between the two lines decreases and the overall 
width of the absorption feature become narrower. 
The double-horn feature in our plots are similar to those described by
Wang (1995) except that Wang assumed a steady state outflow with lower gas 
density and relatively higher gas temperature resulting in shallower 
absorption lines.

The results for model 0M and -2M can be taken as examples of what we would 
expect if galactic winds were responsible for Ly $\alpha$ clouds at $z=3-4$
and at $z <1$, respectively.  Recent KECK observations of Ly $\alpha$ forests 
by Lu et al.(1998) show us interesting absorption lines
which have double horn profile:
two absorption systems at $z_a=2.95614$ in the spectrum of 
QSO1107+4847 and at $z_a=3.62361$ in the spectrum of QSO1422+2309. 
These profiles are very much like one shown in Figure \ref{fig-13}.
They are associated with CIV absorption lines
and located in the visinity of quasars within $3000 \rm km \ s^{-1}$.
If there are nearby galaxies in a cluster around the quasars,
the profiles can be explained as consequence of metal-enriched galactic 
outflows.


Ciardi \& Ferrara (1997) have proposed that a hot secondary halo that
forms  when the inner boundary of the shell re-expands back into the cavity
would also produce Ly $\alpha$ absorption lines.  The beginnings of
this inward re-expansion is seen in our simulation of model 0M.  The outer
radius expanded out to the radius at which we stopped all our simulations
well before the secondary halo had been established. We were, therefore,
not able to explore its effects on quasar spectra.  

Finally, the fact that outflow from low-mass galaxies can extend out to 
distances of 40 kpc or more indicates that such galaxies may have played an
important role in polluting the intergalactic medium with metals at
high redshifts.  This possibility was first discussed by 
Silk, Wyse \& Shields (1987).  The ubiquity of the carbon features
observed in quasar absorption systems at $z\approx 3$ (Cowie et al.~1995;
Tytler et al.~1995; Songaila \& Cowie 1996)
implies that metals were dispersed over a large
region, if not uniformly, fairly early in the history of the Universe.  
The fact that the generally accepted hierarchical clustering models for 
structure formation suggest that at high redshifts, the Universe
was dominated by halos with shallow potential wells and, as we have
shown, the fact that the winds from these wells can spread out over large 
distances supports the scenario where early generations of dwarf galaxies
are responsible for polluting the intergalactic medium.  

%

\section{Summary}

In this paper, we have sought to study the interaction between 
supernova-powered gas outflows from low-mass galaxies and the local 
intergalactic medium (IGM).  
We find that even if the supernova explosions are, in principle, able 
to expel the gas from a low-mass galaxy, they may not be able to do so
if the thermal pressure of the local intergalactic medium is sufficiently
high.   The thermal pressure of the IGM resists the outflow, eventually
bringing it to a halt.  The confinement radius, the radius to which 
the gas can expand before being brought to a halt, depends on the IGM
pressure as $R_{\rm max}\propto P_{\rm IGM}^{-0.36}$.  The higher the 
IGM pressure, the smaller the confinement radius.  
We find that the thermal 
pressure of the hot intracluster medium in clusters of galaxies,
for example, is large enough to prevent the gas from expanding much 
beyond the galaxy.  

The interface between the outflow and quiescent IGM is demarcated by
a dense expanding shell formed by the gas swept-up by the outflow. Once
halted, the IGM pressure pushes the shell back into the galaxy.
The collapsing shell is susceptible to Rayleigh-Taylor instability resulting 
in non-uniform collapse of the gas.  The instability enhances small 
perturbations in the shell, causes it to deform into tentacle-like 
structures and eventually fragments into small clouds.

In high density, high temperature regions, the dwarf galaxies are also
likely to be moving at high velocities and therefore, subject to the
effects of ram pressure.  When ram pressure is comparable or higher
than the ambient thermal pressure, it can distort and fragment the shell 
into high density clouds that are then dragged away from the galaxy and
carried downstream.  These high density clouds are potential sites of 
star formation and the spatial distribution of stars newly born in these
clouds will trace out a diffuse tail-like structure. We speculate that 
comet-like galaxies with diffuse tail seen in  z=1.15 cluster  3C324
are such galaxies.  The structure exhibited by these galaxies is
temporary; it will dissolve away as the stars stream away and become
part of a diffuse population of stars in the intracluster environment.

In contrast, the relatively unhindered outflows in low density, low 
temperature environments can drive the shells of swept-up gas out to distances
of 40 kpc or more from the galaxy.  Such shells, if they intersect a quasar
line-of-sight, would give rise to Ly $\alpha$ absorption lines of the kind 
seen in quasar spectra.   Assuming that the Universe is permeated by
a metagalactic UV flux, the absorption features correspond to
HI column densities typically of order $10^{15}$ cm$^{-3}$.  At small
impact parameters, the velocity field of the expanding shell gives rise
to  double-horned absorption profiles.  This feature weakens and disappears
as the  impact parameter increases.  
Finally, the fact that outflow from low-mass galaxies can extend out to 
distances of 40 kpc or more indicates that such galaxies may have played an
important role in polluting the intergalactic medium with metals at early
epochs.

\section*{Acknowledgments}

This research was partially carried out at the Canadian Institute for
Theoretical Astrophysics.  We would like to thank to M. Dickinson for 
his helpful comments and discussions. 
We also thank the anonymous referee for helpful comments 
and poignant suggestions. I.M. acknowledges NSERC for a 
fellowship and A.B. acknowledges support from NSERC through an operating
grant.

\vspace{30pt}

\newpage
 
%
\begin{table}
\caption{IGM Conditions} \label{tab-model}
\begin{tabular}{lllccrr}
\hline \hline
 Model & C$^{\dagger}$ & $ \widetilde{P}_{\rm IGM}$ &  $ T_{\rm IGM}$
   & $n_{\rm ex}$ & $V_{\rm flow}$ & $R_{\rm max}$  \\
 \hline 

5P & P & $10^5   $ & $10^7$ & $10^{-2}$ & 0.  & 1.5 \\
4P & P & $10^4   $ & $10^7$ & $10^{-3}$ & 0.  & 3.8 \\
3P & P & $10^3   $ & $10^7$ & $10^{-4}$ & 0.  & 8.8 \\
2P & P & $10^2   $ & $10^7$ & $10^{-5}$ & 0.  &18.0 \\
0P & P & $10^0   $ & $10^4$ & $10^{-4}$ & 0.  & $>40$ \\
\\
5M & M & $10^5   $ & $10^7$ & $10^{-2}$ & 0. & 1.5  \\
4M & M & $10^4   $ & $10^7$ & $10^{-3}$ & 0. & 3.6  \\
3M & M & $10^3   $ & $10^7$ & $10^{-4}$ & 0. & 8.6 \\
2M & M & $10^2   $ & $10^7$ & $10^{-5}$ & 0. & 18.0  \\
0M & M  & $10^0   $ & $10^4$ & $10^{-4}$ & 0. & $>40$ \\
-2M & M  & $10^{-2}$ & $10^4$ & $10^{-6}$ & 0. & $>40$  \\
\\
5PW & P & $10^5   $ & $10^7$ & $10^{-2}$ & 1000. \\
4PW & P & $10^4   $ & $10^7$ & $10^{-3}$ & 1000. \\
3PW & P & $10^3   $ & $10^7$ & $10^{-4}$ & 1000. \\
\\
5MW & M      & $10^5   $ & $10^7$ & $10^{-2}$ & 1000. \\
4MW & M      & $10^4   $ & $10^7$ & $10^{-3}$ & 1000. \\
3MW & M      & $10^3   $ & $10^7$ & $10^{-4}$ & 1000. \\
3MW6 & M      & $10^3   $ & $10^7$ & $10^{-4}$ &  600. \\
2MW4 & M      & $10^2   $ & $10^7$ & $10^{-5}$ &  400. \\
\\ \hline
\end{tabular}

\medskip

Units are $\rm K cm^{-3}$ for $ \widetilde{P}_{\rm IGM}$;
K for  $ T_{\rm IGM}$;
$\rm  cm^{-3}$ for $n_{\rm ex}$; 
km s$^{-1}$ for $V_{\rm flow}$;
and kpc for $R_{\rm max}$. \\
$^{\dagger}$ Cooling function: P for
 primordial cooling and M for
metal cooling.
\end{table}



\begin{figure*}
\vspace{3cm}
\centerline{\hbox{
\mbox{ ( murakami-fig1.gif ) }
}}
\caption{
Snapshots of isodensity contour and velocity field in r-z plane for model -2M.
Density profiles ($g \ cm^{-3}$) for the cross section at r=0kpc (solid line)
and 2kpc (dotted line) are also shown.
The time measure from the beginning of the simulation is shown at upper right
corner and 3.8E7yr means $3.8 \times 10^7$ yr.
The contour level is set as $\Delta \log \rho = 0.25$.
\label{fig-1}
}
\end{figure*}

\begin{figure*}
\centerline{\hbox{
\vspace{1cm}
\mbox{ ( murakami-fig2.gif ) }
}}
\caption{
The same as Figure 1 but for model 0M.
\label{fig-2}
}
\end{figure*}

\begin{figure*}
\centerline{\hbox{
\vspace{1cm}
\mbox{ ( murakami-fig3.gif ) }
}}
\caption{
The same as Figure 1 but for model 3M.
\label{fig-3}
}
\end{figure*}

\begin{figure*}
\centerline{\hbox{
\vspace{1cm}
\mbox{ ( murakami-fig4.gif ) }
}}
\caption{
The same as Figure 1 but for model 4M.
\label{fig-4}
}
\end{figure*}

\begin{figure*}
\centerline{\hbox{
\psfig{figure=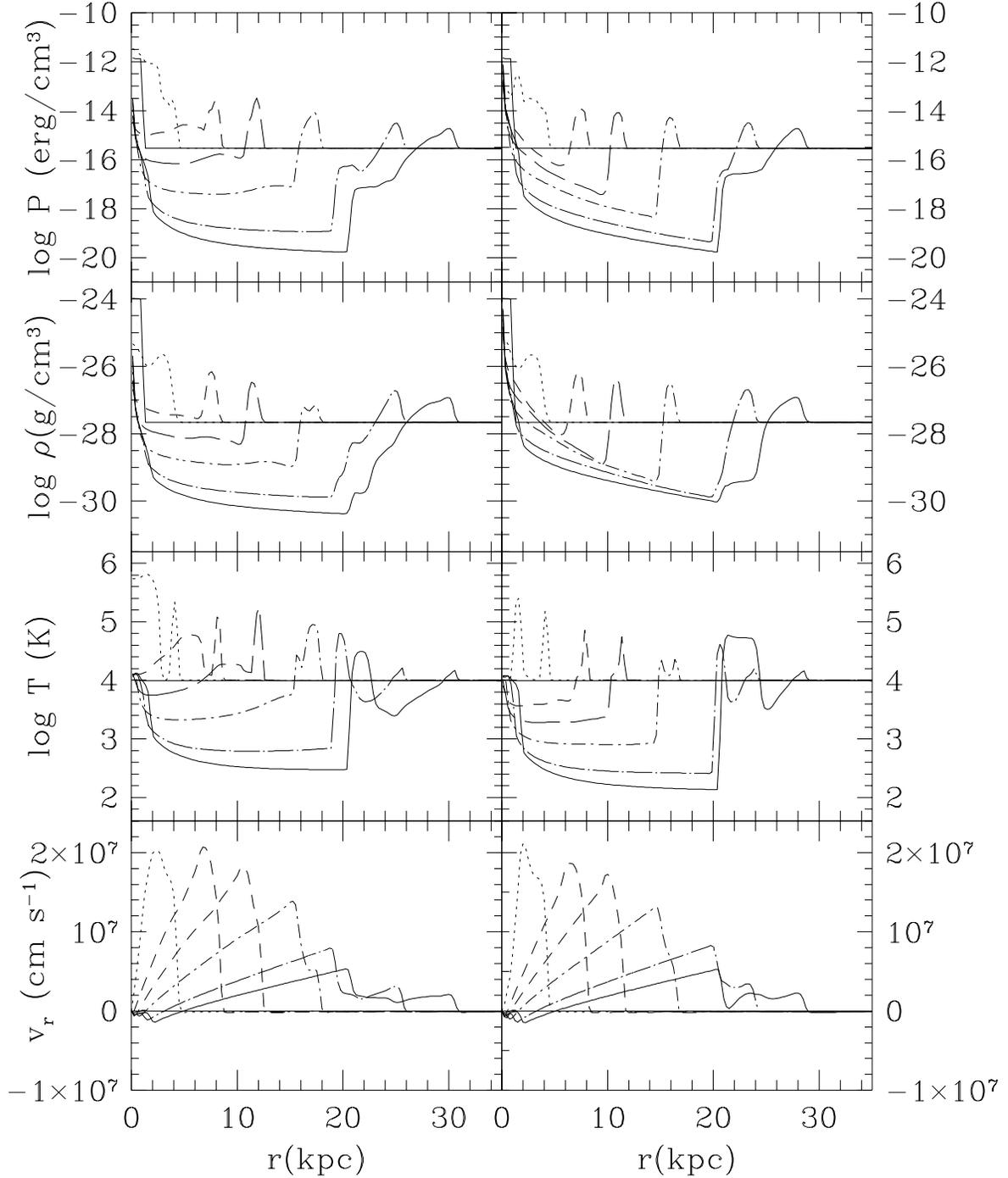,height=22cm,width=16.5cm}
}}
\caption{
Profiles of pressure, density, temperature and velocity (r component)
at the cross section at z=0kpc for model 0P (left) and 0M (right).
The time for each line is initial state (solid line),
$2.4 \times 10^7$ yr (dotted), $4.8 \times 10^7$ yr (dashed),
$7.2 \times 10^7$ yr (long dashed),
$1.2 \times 10^8$ yr (dot-dashed),
$2.4 \times 10^8$ yr (long dash-dotted), and
$3.6 \times 10^8$ yr (solid).
\label{fig-5}
}
\end{figure*}

\begin{figure*}
\centerline{\hbox{
\psfig{figure=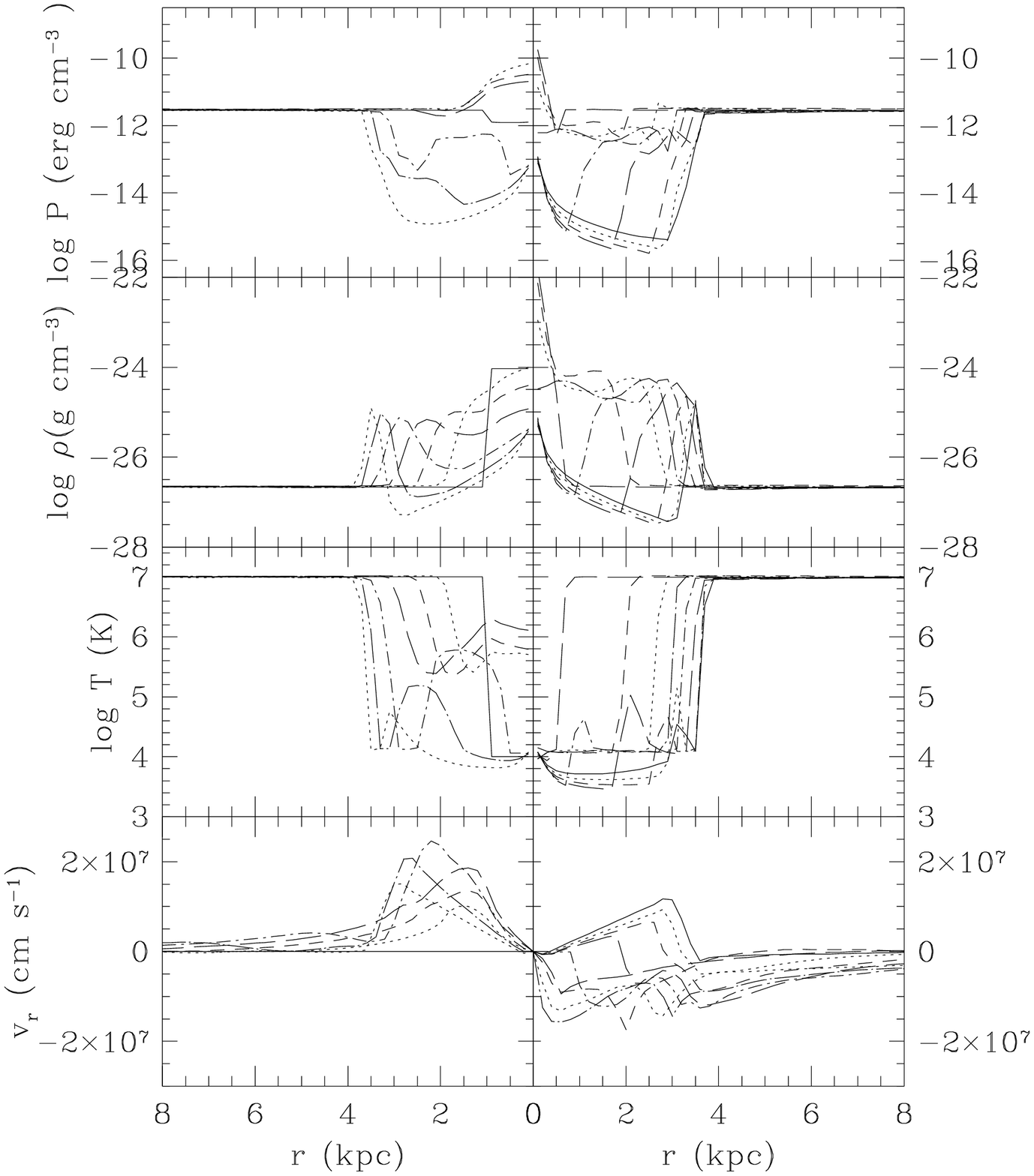,height=22cm,width=16.5cm}
}}
\caption{
Profiles of pressure, density, temperature and velocity (r component)
at the cross section at z=0kpc for model 4M expanding phase(left) 
and collapsing phase (right).
The time for each line is initial state (solid line, left panel),
$9.5 \times 10^6$ yr (dotted, left), 
$1.4 \times 10^7$ yr (dashed, left),
$1.9 \times 10^7$ yr (long dashed, left), 
$2.4 \times 10^7$ yr (dot-dashed, left),
$2.9 \times 10^7$ yr (long dash-dotted, left), 
$3.3 \times 10^7$ yr (dotted, left),
$3.8 \times 10^7$ yr (solid, right),
$4.3 \times 10^6$ yr (dotted, right), 
$4.7 \times 10^7$ yr (dashed, right),
$5.2 \times 10^7$ yr (long dashed, right), 
$5.7 \times 10^7$ yr (dot-dashed, right),
$6.2 \times 10^7$ yr (long dash-dotted, right), 
$6.6 \times 10^7$ yr (dotted, right),
$7.6 \times 10^7$ yr (dashed, right), and
$9.5 \times 10^7$ yr (long dashed, right).
\label{fig-6}
}
\end{figure*}
 
\begin{figure*}
\centerline{\hbox{
\psfig{figure=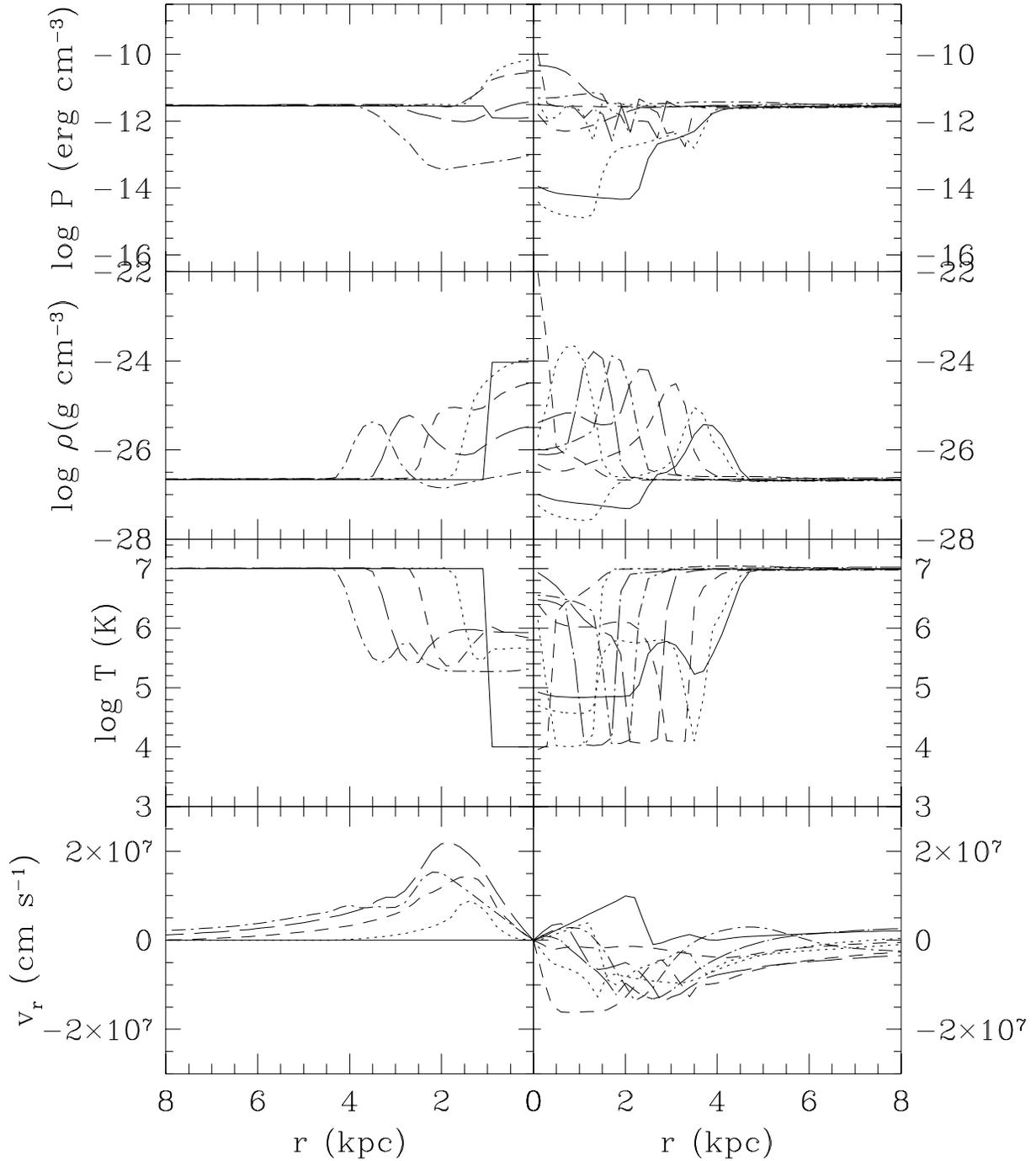,height=22cm,width=16.5cm}
 }}
\caption{
Profiles of pressure, density, temperature and velocity (r component)
at the cross section at z=0kpc for model 4P expanding phase(left) 
and collapsing phase (right).
The time for each line is initial state (solid line, left panel),
$7.6 \times 10^6$ yr (dotted, left), 
$1.5 \times 10^7$ yr (dashed, left),
$2.3 \times 10^7$ yr (long dashed, left), 
$3.0 \times 10^7$ yr (dot-dashed, left),
$3.8 \times 10^7$ yr (solid, right),
$4.6 \times 10^6$ yr (dotted, right), 
$5.3 \times 10^7$ yr (dashed, right),
$6.1 \times 10^7$ yr (long dashed, right), 
$6.8 \times 10^7$ yr (dot-dashed, right),
$7.6 \times 10^7$ yr (long dash-dotted, right), 
$8.3 \times 10^7$ yr (dotted, right), and
$1.1 \times 10^8$ yr (dashed, right). 
\label{fig-7}
 }
\end{figure*}

\begin{figure*}
\vspace{3cm}
\centerline{\hbox{
\mbox{ ( murakami-fig8.gif ) }
}}
\caption{
Snap shots of isodensity contour map  in r-z plane
for model 3M in collapse phase.
The contour level is set as $\Delta \log \rho = 0.25$.
(a) and (b) are results from calculation with numerical resolution, dr=0.2kpc;
and (c) and (d) are results from calculation with dr=0.1kpc.
Density maps of finer numerical resolution show
thinner collapsing shell and many   
tentacle-like features
due to the instability.
1.04E8yr means $ 1.04 \times 10^8$ yr.
\label{fig-8}
}
\end{figure*}

\begin{figure*}
\vspace{2cm}
\centerline{\hbox{
\mbox{ ( murakami-fig9.gif )  }
}}
\caption{
Snapshots of isodensity contour and velocity field in r-z plane for model 3MW.
Density profiles ($g \ cm^{-3}$) for the cross section at r=0kpc (solid line),
2kpc (dotted line), and 5kpc (dashed line) are also shown.
The time measure from the beginning of the simulation is shown at upper right
corner and 4.3E7yr means $4.3 \times 10^7$ yr.
The contour level is set as $\Delta \log \rho = 0.25$.
\label{fig-9}
}
\end{figure*}
 
\begin{figure*}
\vspace{2cm}
\centerline{\hbox{
\mbox{ ( murakami-fig10.gif ) }
}}
\caption{
Snapshots of isodensity contour and velocity field in r-z plane for model 5MW.
Density profiles ($g \ cm^{-3}$) for the cross section at r=0kpc (solid line)
and 1kpc (dotted line) are also shown.
The contour level is set as $\Delta \log \rho = 0.25$.
\label{fig-10}
}
\end{figure*}

\begin{figure}
\centerline{\hbox{
\psfig{figure=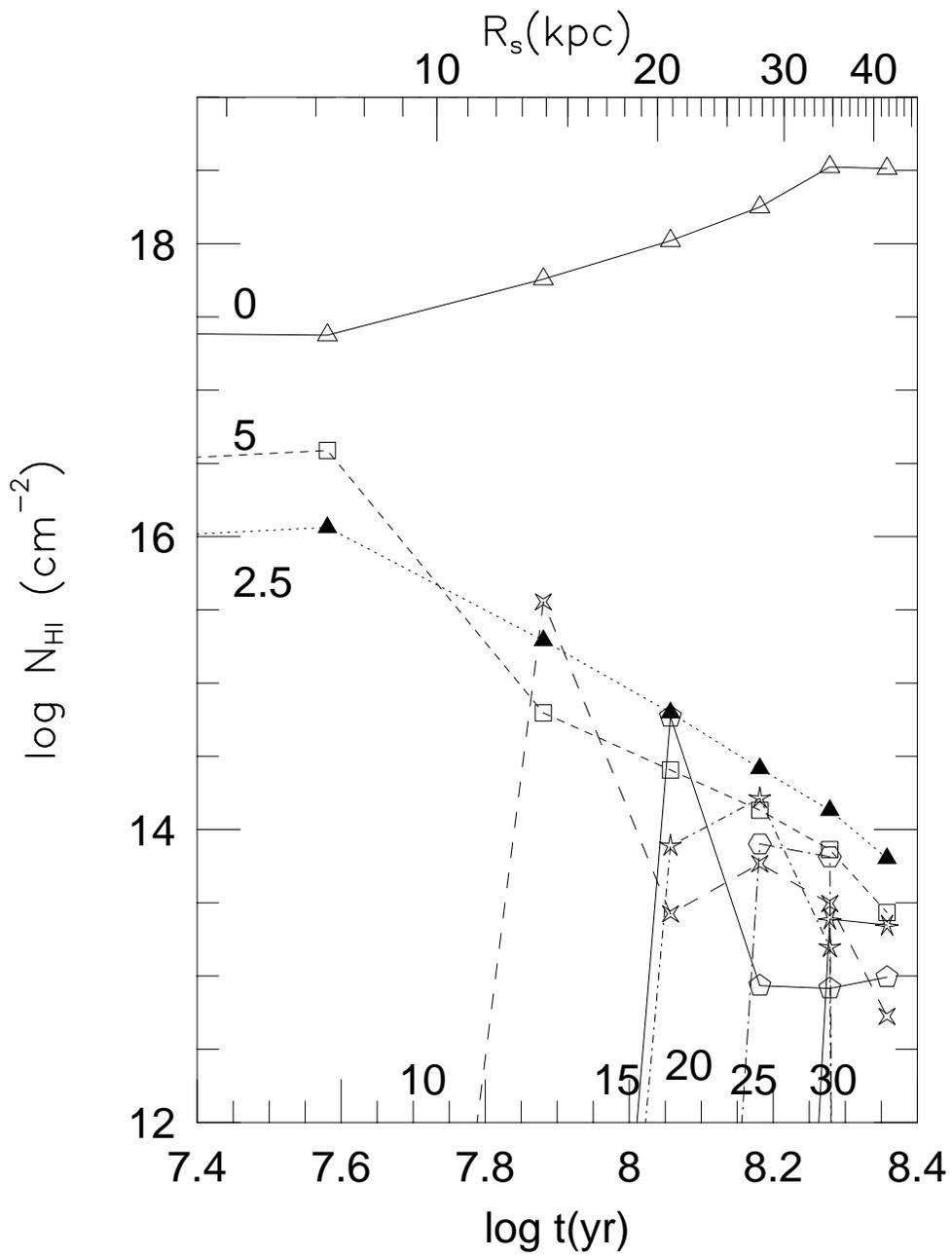,width=16cm}
}}
\caption{
Time evolution of HI column densities for fixed sight lines
for mode -2M.
$J_0=1.0 \times 10^{-21} \rm erg \ s^{-1} cm^{-2} Hz^{-1} str^{-1} $ 
is assumed for the background UV flux.
Each line is labeled with the impact parameters. 
The contribution of the IGM to the column densities is neglected.
The radius of the expanding shell is indicated as the upper horizontal axis.
\label{fig-11}
}
\end{figure}

\begin{figure}
\centerline{\hbox{
\psfig{figure=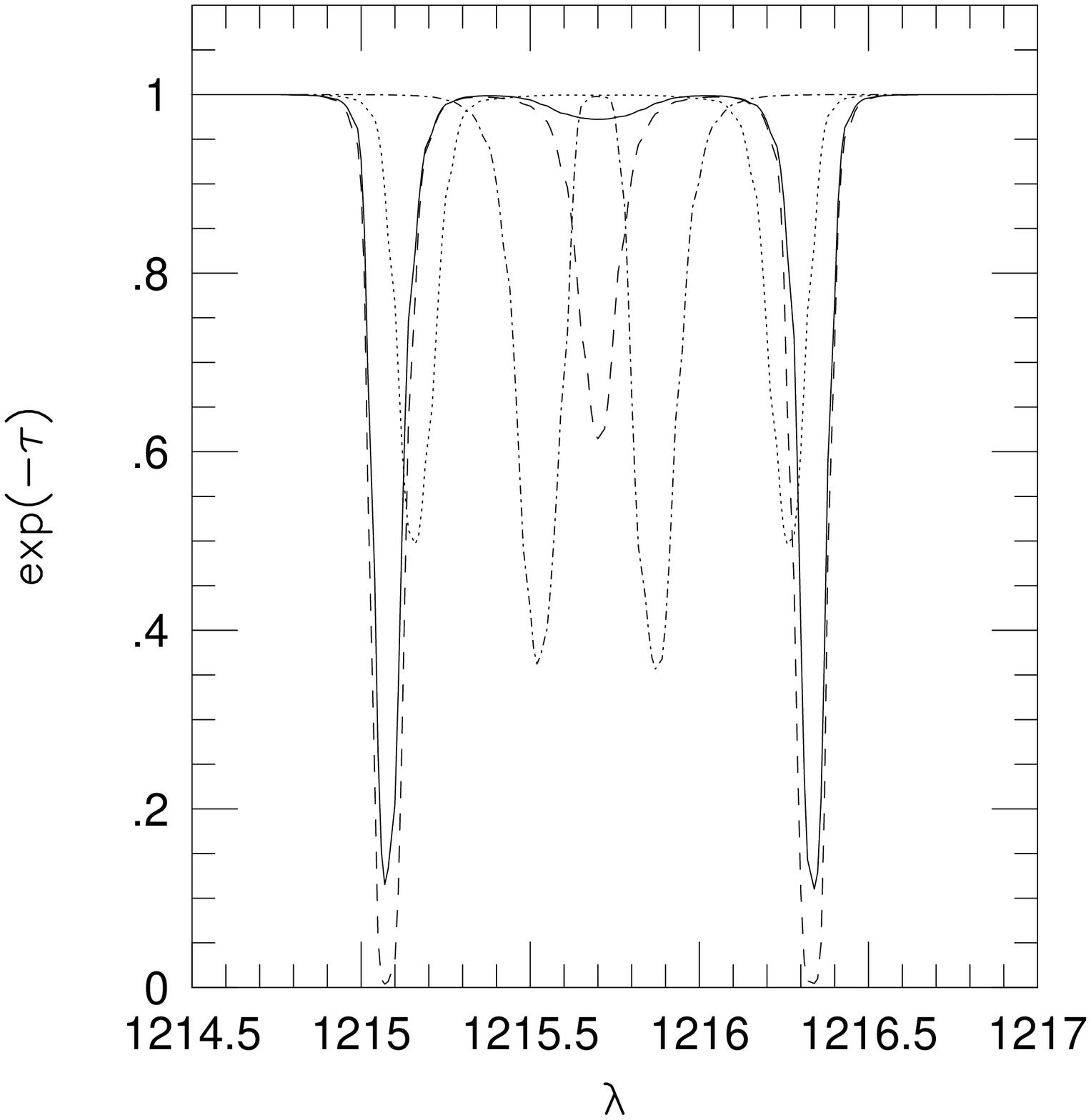,height=20cm,angle=360}
}}
\caption{
Expected absorption profiles for model -2M 
at t=1.9 $\times 10^8$yr.  
Impact parameters of sight lines are 
5kpc (dashed line),
10kpc (solid line), 20kpc (dotted line), and 30kpc (dot-dashed line).
$J_0=1.0 \times 10^{-21} \rm erg \ s^{-1} cm^{-2} Hz^{-1} str^{-1} $ 
is assumed for the background UV flux.
\label{fig-12}
}
\end{figure}

\begin{figure}
\centerline{\hbox{
\psfig{figure=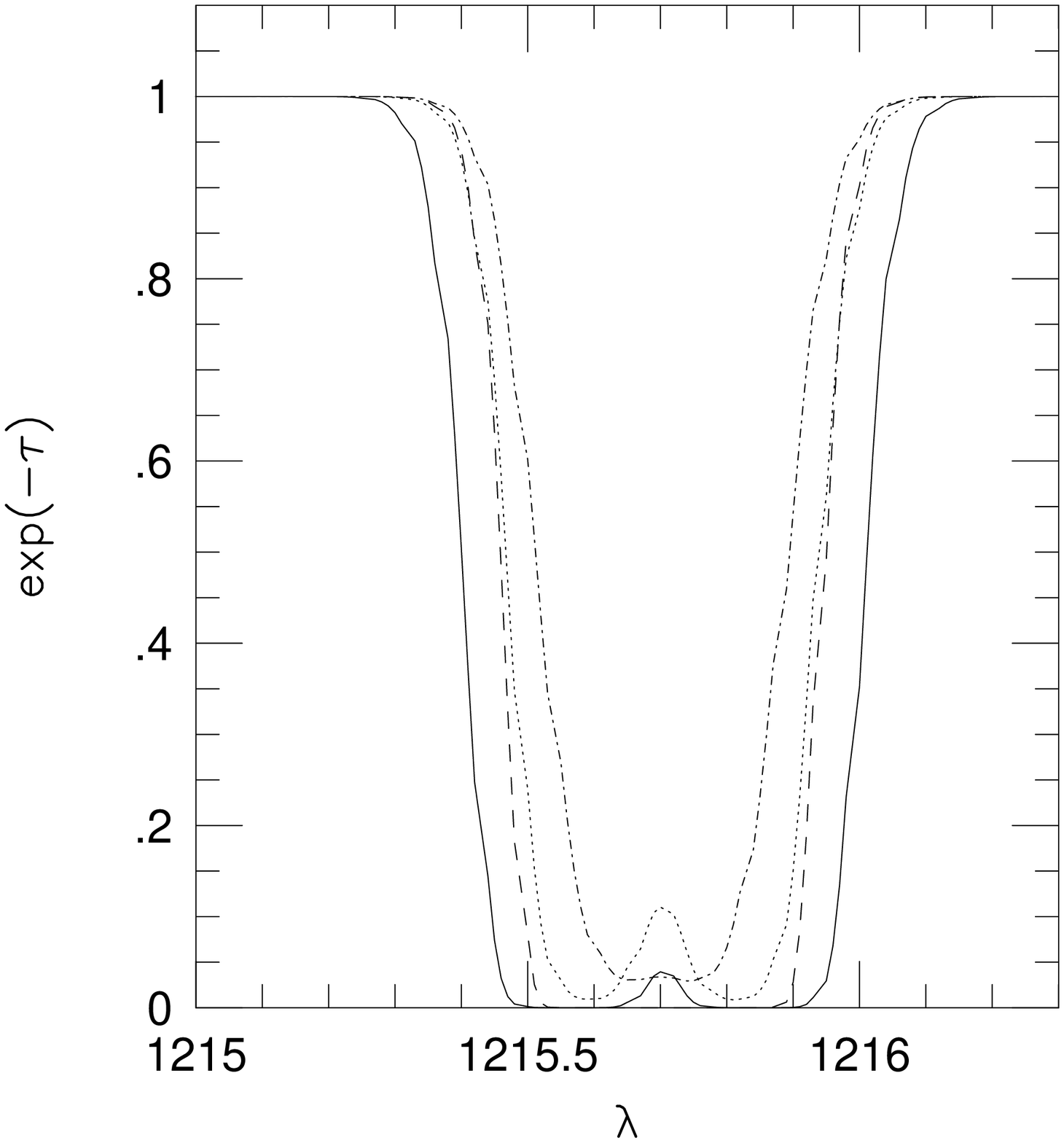,height=20cm,angle=360}
}}
\caption{
Expected absorption profiles for model 0M 
at t=2.9 $\times 10^8$yr.  
Impact parameters of sight lines are 
10kpc (solid line), 15kpc (dotted line), 20kpc (dot-dashed line),
and 25kpc(dashed line). 
$J_0=1.0 \times 10^{-21} \rm erg \ s^{-1} cm^{-2} Hz^{-1} sr^{-1} $ 
is assumed for the background UV flux.
\label{fig-13}
}
\end{figure}


\begin{thebibliography}{}

\bibitem[\protect\citename{Allen }1973]{allen}
Allen C. W., 1973, in {\it Astrophysical Quantities}, 3rd edition. Athlone 
Press, London.

\bibitem[\protect\citename{Arnet et al.\ }1991]{afm}
Arnett  D., Fryxell  B., Muller  E., 1991, in {\it Supernovae}, Ed:
S. E. Woosley, Springer, New York, p. 232

\bibitem[\protect\citename{Babul \& Ferguson }1996]{bf}
Babul A., Ferguson H. C., 1996, ApJ, 458, 100

\bibitem[\protect\citename{Babul \& Rees }1992]{br}
Babul A., Rees M. J., 1992, MNRAS, 255, 346 


\bibitem[\protect\citename{Balsara et al.\ }1994]{blod}
Balsara D. S., Livio M.,  O'Dea C. P., 1994, ApJ, 437, 83



\bibitem[\protect\citename{Black }1981]{black}
Black J. H., 1981, MNRAS, 197, 553


\bibitem[\protect\citename{Castor et al.\ }1975]{cmw}
Castor J., McCray R., Weaver R.,  1975, ApJ, 200, L103 

\bibitem[\protect\citename{Campos }1997]{campos}
Campos A., 1997, ApJ, 488, 606

\bibitem[\protect\citename{Chebalier }1976]{chebalier}
Chevalier R. A., 1976, ApJ, 207, 872

\bibitem[\protect\citename{Ciardi \& Ferrara }1997]{cf}
Ciardi B., Ferrara A., 1997, ApJ, 483, L5 

\bibitem[\protect\citename{Corbelli et al.\ }1997]{cgp}
Corbelli E., Galli D., Palla F., 1997 ApJ, 487, L53

\bibitem[\protect\citename{Couch et al.\ }1994]{cess}
Couch W. J., Ellis R. S., Sharples R. M., Smail I., 1994, ApJ, 430, 121

\bibitem[\protect\citename{Couch et al.\ }1998]{cbses}
Couch W. J., Barger A., Smail I.,  Ellis R. S., Sharples R. M., 1998, ApJ, 
497, 188

\bibitem[\protect\citename{Cowie et al.\ }1995]{cowie95}
Cowie L.L., Songaila A., Kim T.-S., \& Hu E.M., 1995, AJ, 109, 1522

\bibitem[\protect\citename{De Young \& Gallagher }1990]{dyg}
De Young D. S., Gallagher J. S. III, 1990, ApJ, 356, L15

\bibitem[\protect\citename{De Young \& Heckman }1994]{dyh}
De Young D. S., Heckman T., 1994, ApJ, 431, 598

\bibitem[\protect\citename{Dekel \& Silk }1986]{ds}
Dekel  A., Silk  J., 1986, ApJ, 303, 39

\bibitem[\protect\citename{Dickinson }1996]{dickinson}
Dickinson M., 1996, in {\it HST and the High Redshift Universe}, Eds:
N. Tanvir, A. Aragon-Salamanca, J.V. Wall, World Scientific, London. 
(astro-ph/9612178)

\bibitem[\protect\citename{Dressler et al.\ }1994]{dobg}
Dressler A., Oemler A., Butcher H. R., Gunn J. E., 1994, ApJ, 430, 107


\bibitem[\protect\citename{Driver \& Fernandez-Soto }1998]{drs}
Driver S. P.,  Fernandez-Soto A., 1998, 
in {\it Dwarf Galaxies and Cosmology}, 
Eds: Thuan et. al.  Editions Frontiers. 
(astro-ph/9805100).

\bibitem[\protect\citename{Efstathiou }1992]{efstathiou92}
Efstathiou, G., 1992, MNRAS, 256, 43P

\bibitem[\protect\citename{Fabian et al.\ }1980]{fsf}
Fabian A. C., Schwarz J.,  Forman W., 1980, MNRAS, 192, 135

\bibitem[\protect\citename{Ferguson }1992]{ferguson92}
Ferguson H. C., 1992, MNRAS, 255, 389


\bibitem[\protect\citename{Ferguson \& Binggeli }1994]{fb}
Ferguson H. C., Binggeli B., 1994, A\&AR, 6, 67

\bibitem[\protect\citename{Ferguson \& Sandage }1991]{fs}
Ferguson H. C., Sandage A., 1991, AJ, 101, 765

\bibitem[\protect\citename{Ferguson et al.\ }1998]{ftvh}
Ferguson H. C., Tanvir N. R., Von Hippel T., 1998, Nature, 391, 461
(astro-ph/9801228).

\bibitem[\protect\citename{Fioc \& Rocca-Volmerange }1998]{frv}
Fioc M., Rocca-Volmerange B., 1999, astro-ph/9902020

\bibitem[\protect\citename{Forcada-Mir\'{o} }1997]{forcada}
Forcada-Mir\'{o} M. I., 1997, astro-ph/9712205

\bibitem[\protect\citename{Gaetz et al.\ }1987]{gss}
Gaetz T. J., Salpeter E. E.,  Shaviv G., 1987 ApJ, 316, 530

\bibitem[\protect\citename{Gisler }1976]{gisler}
Gisler G. R., 1976, A\&A, 51, 137

\bibitem[\protect\citename{Gunn \& Gott }1972]{gg}
Gunn J. E.,  Gott J. R., 1972, ApJ, 176, 1

\bibitem[\protect\citename{Hachisu et al.\  }1991]{hmns}
Hachisu I., Matsuda T., Nomoto K.,  Shigeyama T., 1991, ApJ, 368, L27 

\bibitem[\protect\citename{Heckman et al.\ }1995]{heckman}
Heckman T.M., Dahlem M., Lehnert M.D., Fabbiano G., Gilmore D., Waller W. H.,
 1995, ApJ, 448, 98

\bibitem[\protect\citename{Ikeuchi }1986]{ikeuchi}
Ikeuchi S., 1986, Astrophys. Space Sci., 118, 509

\bibitem[\protect\citename{Katz, Weinberg \& Hernquist }1996]{kwh}
Katz N., Weinberg D. H., Hernquist L., 1996, ApJS, 105, 19


\bibitem[\protect\citename{Kepner et al.\ }1997]{kbs}
Kepner J., Babul A., Spergel D., 1997, ApJ, 487, 61

\bibitem[\protect\citename{Koo et. al.\  }1997]{koo}
Koo D. C., Guzman R., Gallego J., Wirth G. D., 1997, ApJ, 478, 49


\bibitem[\protect\citename{Larson }1974]{la74}
Larson R. B., 1974, MNRAS, 166, 585

\bibitem[\protect\citename{Larson }1987]{la87}
Larson R. B., 1987, in {\it Starbursts and Galaxy Evolution}, Eds:
T. Thuan, T. Montmerle, J. Tranthanhvan,  Edition Frontieres, p467

\bibitem[\protect\citename{Lea \& De Young }1976]{ldy}
Lea S. M.,  De Young D. S., 1976, ApJ, 210, 647

\bibitem[\protect\citename{Lehnert \& Heckman }1996]{lh}
Lehnert M., Heckman T., 1996, ApJ, 472, 546

\bibitem[\protect\citename{Lu et al.\ }1998]{lu}
Lu L., Sargent W. L. W., Barlow T. A., Rauch M., 1998, astro-ph/9802189

\bibitem[\protect\citename{MacLow \& Ferrara} 1998]{mf}
MacLow M.-M., Ferrara, A., 1998, in {\it The Magellanic Clouds and Other 
Dwarf Galaxies}, Eds.: T. Richtler and J.M. Braun, Shaker Verlag, Aachen.
(astro-ph/9801237)


\bibitem[\protect\citename{MacLow et al.\ }1989]{mmn}
MacLow M.-M., McCray R., Norman M. L., 1989, ApJ, 337, 141

\bibitem[\protect\citename{Mathews }1972]{mathews}
Mathews W., 1972, ApJ, 174, 101 

\bibitem[\protect\citename{Marlowe et al.\ }1995]{marlowe}
Marlowe A. T., Heckman T. M., Wyse R. F. G., Schommer, R.,
1995, ApJ, 438, 563

\bibitem[\protect\citename{Meurer et al.\ }1992]{meurer}
Meurer G. R., Freeman K. C., Dopita M. A., Cacciari C.,
1992, AJ, 103, 60 

\bibitem[\protect\citename{Meurer et al.\ }1996]{moore1}
Moore B., Katz N., Lake G., Dressler A., 1996, Nature, 379, 613

\bibitem[\protect\citename{Meurer et al.\ }1998]{moore2}
Moore B., Lake G., Katz N., 1998, ApJ, 495, 139

\bibitem[\protect\citename{Murakami \& Ikeuchi }1994]{mi}
Murakami I.,  Ikeuchi S.,  1994, ApJ, 420, 68


\bibitem[\protect\citename{Nagasawa et al.\ }1988]{nnm}
Nagasawa M., Nakamura T.,  Miyama S. M., 1988, PASJ, 40, 691

\bibitem[\protect\citename{Navarro \ Steinmetz }1997]{ns97}
Navarro J. F., Steinmetz M., 1997, ApJ, 478, 13


\bibitem[\protect\citename{Norman \& Spaans }1997]{ns}
Norman C., Spaans M., 1997, ApJ, 480, 145

\bibitem[\protect\citename{Norman \& Winkler }1986]{nw}
Norman M. L., Winkler K-H. A., 1986, in {\it Astrophysical Radiation 
Hydrodynamics}, Eds: K-H. A. Winkler, M. L. Norman,  Dordrecht, Reidel, p. 187

\bibitem[\protect\citename{Nulsen}1982]{nul}
Nulsen P. E. J., 1982, MNRAS, 198, 1007

\bibitem[\protect\citename{Peterson \& Caldwell }1993]{pc}
Peterson R. C., Caldwell N., 1993, AJ, 105, 1411

\bibitem[\protect\citename{Portnoy et al.\ }1993]{pps}
Portnoy D., Pistinner S.,  Shaviv G., 1993, ApJS, 86, 95

\bibitem[\protect\citename{Quinn et al.\ }1996]{quinn}
Quinn T., Katz N., Efstathiou G., 1996, MNRAS, 278, L49

\bibitem[\protect\citename{Raymond et al.\ }1976]{rcs}
Raymond J. C., Cox D. P.,  Smith B. W., 1976, ApJ, 204, 290

\bibitem[\protect\citename{Rees }1986]{rees}
Rees M. J., 1986, MNRAS, 218, 25p

\bibitem[\protect\citename{Rieke et al.\ }1993]{rlrt}
Rieke G., Loken L., Rieke M., Tamblyn P. 1993, ApJ, 214, 99


\bibitem[\protect\citename{Saito }1979]{saito}
Saito M., 1979, PASJ, 31, 193

\bibitem[\protect\citename{Secker }1996]{secker}
Secker J., 1996, ApJ, 469, L81

\bibitem[\protect\citename{Secker et al. }1997]{secker97}
Secker J., Harris W.E.,  Plummer J. D., 1997, PASP, 10, 1377

\bibitem[\protect\citename{Silk, Wyse \& Shields }1987]{sws}
Silk  J., Wyse, R.F.G., Shields, G.A., 1987, ApJ, 322, L59

\bibitem[\protect\citename{Songaila \& Cowie }1996]{sc}
Songaila A.,  Cowie L.L., 1996, AJ, 112, 335

\bibitem[\protect\citename{Spaans \& Norman }1997]{sn}
Spaans M., Norman C., 1997, ApJ, 483, 87

\bibitem[\protect\citename{Takeda et al.\ }1984]{tnf}
Takeda H., Nulsen P. E. J.,  Fabian A. C., 1984 MNRAS, 208, 279

\bibitem[\protect\citename{Theuns \& Warren }1997]{tw}
Theuns T., Warren S. J., 1997, MNRAS, 284, L11

\bibitem[\protect\citename{Thoul \& Weinberg }1996]{thoul}
Thoul A. A., \& Weinberg D. H., 1996, ApJ, 465, 608

\bibitem[\protect\citename{Thuan et al.\ }1991]{thuan}
Thuan T. X., Alimi J.-R., Gott J. R. III, Schneider S. E., 
1991, ApJ, 370, 25

\bibitem[\protect\citename{Tolstoy }1999]{tolstoy99}
Tolstoy E. 1999, astro-ph/9901245

\bibitem[\protect\citename{Toyama \& Ikeuchi }1980]{ti}
Toyama K., Ikeuchi S., 1980, Progress Theor. Phys. 64, 831

\bibitem[\protect\citename{Trentham }1998]{tr}
Trentham N., 1998, MNRAS, 294, 193

\bibitem[\protect\citename{Tytler }1995]{tytler}
Tytler D., Fan X.-M., Burles S., Cottrell L., Davis C., Kirkman D.,
Zuo L., 1995, in {\it QSO Absorption Lines}, Ed: G. Meylan,
Springer-Verlag.

\bibitem[\protect\citename{Umemura \& Ikeuchi }1984]{ui}
Umemura M., Ikeuchi S., 1984, Prog. Theor. Phys., 72, 47

\bibitem[\protect\citename{Vader }1986]{vader}
Vader J. P., 1986, ApJ, 305, 669

\bibitem[\protect\citename{Vader \& Chaboyer }1993]{vc}
Vader J. P.,  Chaboyer B., 1993, AJ, 108, 1209

\bibitem[\protect\citename{Vader \& Sandage }1991]{vs}
Vader J. P., Sandage A., 1991,  ApJ, 379, L1

\bibitem[\protect\citename{Van den Bergh }1994]{vdb94}
Van den Bergh S., 1994, ApJ, 428, 617

\bibitem[\protect\citename{Wang }1995]{wang}
Wang B., 1995, ApJ, 444, L17

\bibitem[\protect\citename{White \& Frenk }1991]{wf}
White S. D. W., Frenk C. S., 1991, ApJ, 379, 52


\bibitem[\protect\citename{Yoshioka \& Ikeuchi }1990]{yi}
Yoshioka S.,  Ikeuchi S., 1990, ApJ, 360, 352


\end{thebibliography}
\end{document}